\documentclass{article}
\usepackage{graphicx}
\usepackage{epsfig}
\usepackage{amsfonts}
 \usepackage{amssymb}
\usepackage[dvips]{color}
\date{\today}

\newcommand{\insertplot}[5]{\begin{figure}
 \hfill\hbox to 0.05in{\vbox to #5in{\vfill
 \inputplot{#1}{#4}{#5}}\hfill}
 \hfill\vspace{-.1in}
 \caption{#2}\label{#3}
 \end{figure}}
 \newcommand{\inputplot}[3]{
 \special{ps: plotfile #1}
}
\newcounter{fig}

\newcommand{\ee}{\end{equation}}
\newcommand{\eea}{\end{eqnarray}}
\newcommand{\be}{\begin{equation}}
\newcommand{\bea}{\begin{eqnarray}}

\begin{document}

 \title{Boson stars and black holes with wavy scalar hair} 

\author{
{\large Yves Brihaye}$^{(1)}$, 
{\large } 
and {\large Betti Hartmann$^{(2)}$} 
\\
\\
$^{(1)}${\small  Service de Physique de l'Univers, Champs et Gravitation, Universit\'e de Mons, Mons, Belgium}\\
$^{(2)}${\small  Department of Mathematics, University College London, Gower Street, London, WC1E 6BT, UK }
}
\maketitle 
\begin{abstract} 
In this paper, we follow up on the discovery of a new type of solution in the Einstein-Maxwell system
coupled minimally to a self-interacting complex scalar field. For sufficiently large gravitational
coupling and sufficiently small electromagnetic coupling we demonstrate that 
boson stars as well as black holes can carry scalar hair that shows a distinct new feature~:
a number of spatial oscillations in the scalar field away from the core or horizon, respectively. These spatial
oscillations appear also in the curvature invariants and hence should be a detectable feature of
the space-time. As a first hint that this is true, we show that the effective potential
for null geodesics in this space-time possesses a local minimum indicating that in the spatial region where oscillations
occur a new stable photon sphere should be possible. 
We also study the interior of the black holes with scalar hair and show that the curvature singularity appears at a finite
value of the radius and that black holes with wavy scalar hair have this singularity very close to the center.
 \end{abstract}

\section{Introduction}
In the quest to understand the nature of dark energy, which currently seems to dominate the energy density content of our universe, 
different routes can be taken within the given framework of the Standard Model of Particle Physics and/or our model of the universe
which relies successfully on the Theory of General Relativity (GR). One route to understand how a model describing all known interactions would look like is to first understand how fields that typically appear in the Standard Model of Particle physics behave in strongly curved space-times.

One of the earliest steps in this direction was done by studying electromagnetic fields coupled minimally to General Relativity.  
As in the case of the vacuum solution to the Einstein equation, a spherically symmetric solution to the combined Einstein-Maxwell equations
is static and given by the Reissner-Nordstr\"om metric, which -- in turn -- is a solution uniquely described by its asymptotic ''charges'', i.e. the ADM mass and the electric (and/or magnetic) charge. 
This statement, together with new results for rotating, axially symmetric space-times, has then led to the formulation of the
No-hair conjecture \cite{ruffini}. In fact, for spherically symmetric, asymptotically flat electro-vacuum space-times in the context of GR, this now is a strictly proven (No-hair) theorem.
The status of the conjecture, however, is less clear when changing 
any of the conditions menioned above. As an example, in recent years so-called {\it scalar-tensor} gravity models have been studied extensively, essentially building on the original work by Horndeski \cite{horndeski} (and more recently \cite{Nicolis:2008in}).
In these extensions of General Relativity a scalar field is non-minimally coupled to combinations of curvature tensors, e.g. the Gauss-Bonnet term. This is typically done in such a way that the
resulting equations of motion remain of second order. In these models,
black holes can carry non-trivial scalar fields on the horizon.
However, one does not have to leave the realm of General Relativity to find counterexamples to the conjecture. 
Introducing a complex, massive scalar field into GR,
it has been shown \cite{Herdeiro:2014goa} (see also the review \cite{Herdeiro:2015waa}) that sufficiently fast rotating Kerr black holes that fulfill a precise fine-tuning between
the angular velocity at the horizon and the phase of the scalar field (denominated a {\it synchronisation condition}) can carry scalar hair. 
Interestingly, it was shown in \cite{Herdeiro:2014goa} that these solutions bifurcate from a very specific subset of the Kerr family
as a result of an instability of the latter. 

The question remains whether the rotation of the black hole is 
necessary to form scalar hair, or if non-rotating solutions
can also become unstable in the sense discussed above.
It appears that this is, indeed, possible, but the electromagnetic fields as well as the self-interaction of the scalar field are essential ingredients here. 
The first example of such solutions were presented in \cite{Hong:2019mcj}. In this case, an O(3) scalar field was minimally coupled to GR hence evading the hypothesis of the no-hair theorems given in
\cite{Nunez:1996xv,Hod:2012px,Mayo:1996mv}.
In fact, black holes with scalar hair are also possible 
in a complex scalar field model with the scalar field gauged under a U(1) symmetry. This was studied first in \cite{Herdeiro:2020xmb} (see also \cite{Hong:2020miv} ). Considering first the 
Maxwell-scalar field equations in the background of a Schwarzschild black hole, the authors constructed so-called {\it Q-clouds} surrounding black holes. They found that these clouds of scalar field
exist only if both the electric charge of the cloud as well as the
self-interaction of the scalar field are present. 
These results can be extended to the case in which the space-time becomes dynamical and as an extension to the results in \cite{Herdeiro:2020xmb,Hong:2020miv} it was shown in \cite{Brihaye:2020vce,Brihaye:2020yuv,Brihaye:2021phs} that additional branches of solutions with new features exist. 
In particular, it was shown
that for sufficiently large gravitational coupling, the space-time splits into two distinct parts: (a) an inflating interior and (b) an exterior which is described by the extremal Reissner-Nordstr\"om solution \cite{Brihaye:2020vce}. Moreover, when the electric field on the horizon is sufficiently small, an intermediate region appears in which the scalar field develops spatial oscillations ({\it wavy scalar hair) } which can also be found in the curvature invariants \cite{Brihaye:2021phs}. 
It is exactly this latter feature that we are going to elaborate more on in this paper. As an extension to the results presented in \cite{Brihaye:2021phs}, we will demonstrate here that these oscillations also appear in the space-time of a charged boson star.
Boson stars \cite{kaup,misch,flp,jetzler,new1} are globally regular lumps of scalar field that carry a conserved Noether charge associated to the U(1)
symmetry of the model. When considered in flat space-time, these solutions are often referred to as {\it Q-balls} \cite{kusenko,dm},  the existence of which
might have interesting implications \cite{implications}. 

Next to boson stars we have also constructed the interior of the black hole solutions
and demonstrate that the curvature singularity appears at a finite distance from the origin of the coordinate system. 
Finally, we show that the black holes (and boson stars, respectively) with wavy scalar hair possess stable circular photon orbits on the surface of the scalar cloud. 

\section{Set-up and equations of motion}
The model studied in this paper has been discussed in a number of publications before and we refer the reader to 
these papers for more details (see e.g. \cite{Herdeiro:2020xmb}). For completeness, we would like to remind the reader of the action
as well as the equations of motion that we have solved subject to appropriate boundary conditions.

The action of the model is
\be
\label{eq:action}
{\cal{L}} = \frac{1}{16 \pi G} R   - \frac{1}{4} F^{\mu \nu} F_{\mu \nu} 
- D_{\mu} \Psi^{\dagger}  D_{\mu} \Psi - U(|\Psi|) 
\ee
describing a U(1) gauge field $A_{\mu}$ with field strength tensor 
$F_{\mu \nu}$ and a complex valued, self-interacting scalar field $\Psi$ both interacting minimally
with gravity as well as minimally {\it inter se}. $D_{\mu} = \partial_{\mu} - i g A_{\mu}$ denotes the covariant derivative
and $U(|\Psi|)$ the scalar field potential. The crucial point about the solutions that we are presenting here is that
the scalar field is charged and self-interacting. In the following, we will choose the following potential
 \be 
\label{eq:poly}
 U(\vert\Psi\vert) = \mu^2 \vert\Psi\vert^2 - \lambda \vert\Psi\vert^4 + \nu \vert\Psi\vert^6 \  , 
\ee 
but emphasize that the results in \cite{Brihaye:2021phs} demonstrate that the actual
form of the potential is not that relevant for the qualitative behaviour of the solutions. 
We will discuss spherically symmetric solutions to the equations resulting from
the variation of the action (\ref{eq:action}). While the space-time will be static and allows the Ansatz for the metric in the form~:
\begin{equation}
{\rm d}s^2 = -(\sigma(r))^2 N(r) {\rm d}t^2 + \frac{1}{N(r)} {\rm d}r^2 + r^2\left({\rm d}\theta^2 + \sin^2 \theta {\rm d}\varphi^2 \right)  \ \ , \ \
N(r)=1-\frac{2 \tilde{m}(r)}{r} \ , 
\end{equation}
the matter field Ansatz is stationary
\begin{equation}
A_{\mu} {\rm d} x^{\mu} = v(r) {\rm d} t  \ \ , \ \ 
\Psi=\psi(r) \exp(i\omega t) \ ,
\end{equation} 
however, leads to a static energy-momentum tensor. 
Note also that due to the U(1) gauge symmetry, the field equations depend only on the combination $g v(r) - \omega$.
This allows us to fix the residual gauge freedom by setting $\omega = 0$ in the following. 
The explicit form of the equations of motions to be solved numerically then reads~:
\begin{equation}
    \tilde{m}' = 4 \pi G r^2 \biggl[ \frac{v'^2}{2 \sigma^2} + N \psi'^2 + U(\psi) + \frac{(g v \psi)^2}{N \sigma^2} \biggr]  \ \ , \ \ 
		\sigma' =  8 \pi G r \sigma \biggl[ \psi'^2 + \frac{(g v \psi)^2}{N^2 \sigma^2} \biggr]  \ \ , \ 
\label{eq:einstein}
\end{equation} 
\begin{equation}
		v''
	+  \biggl[ \frac{2}{r} -\frac{\sigma'}{\sigma} - \frac{2 \gamma}{r^2} \left((1-N) \frac{\sigma'}{\sigma} + N' \right) \biggr] v' 
		= \frac{2 g^2 v \psi^2}{N}  \ \ , \ \ 
			\psi'' + \left(\frac{2}{r} + \frac{N'}{N} +\frac{\sigma'}{\sigma}\right) \psi' + \frac{g^2 v^2 \psi}{N^2 \sigma^2} - \frac{1}{2N} \frac{dU}{d \psi} =0 \ \  ,
\label{eq:matter}
\end{equation} 
where the prime denotes the derivative with respect to $r$. 

Using the rescalings $x=\mu r$, $m=\mu \tilde{m}$, $V= \frac{\sqrt{\lambda}}{\mu} v$, $\psi = \frac{\sqrt{\lambda}}{\mu} \Psi$ we are left with three dimensionless couplings~:
\be
      \alpha = \frac{4 \pi G \mu^2}{\lambda} \ \ , \ \ 
			\beta^2 = \frac{\nu \mu^2}{\lambda^2} \ \ , \ \ 
			e = \frac{g}{\sqrt {\lambda}} \ . 
\ee
Note that the potential parameter $\beta^2$ has to be chosen with care in order to find the desired solutions.
For $\beta^2=1/4$ the potential  possesses degenerate vacua at $\psi= 0$,  $\psi^2 = 2$, while for
 $\beta^2=1/3$ the potential possesses a saddle point at $\psi^2 = 1$. In the following, we will choose $\beta^2 = 9/32$, a value
in between these two choices. 

The solutions can be characterized by their (dimensionless) mass $M$ and their (dimensionless) electric charge $Q$ - similar to the spherically
symmetric static solutions with vanishing scalar fields, the Reissner-Nordstr\"om solution. However, the scalar field 
and in particular the internal symmetry associated to it add another physical property, namely the globally conserved Noether charge.
While the former two can be read of from the asymptotic behaviour of the gravitational and electric fields
\begin{equation}
\label{eq:infty}
N(x \gg 1)=1-\frac{2M}{x} + \frac{\alpha Q^2}{x^2} + ..... \ \ , \ \ 
V(x\gg 1)= \Phi  - \frac{Q}{x} + ....
\end{equation}
with $\Phi$ a as constant, 
the latter is given in terms of the integral  of the $t$-component of the locally conserved Noether current~:
\begin{equation}
\label{eq:noether}
Q_N=\int {\rm d} x \ \frac{2x^2 e V\psi^2}{N\sigma}  \ .
\end{equation}
In fact, the solutions with non-trivial scalar fields can be thought of as being surrounded by a ``cloud'' of scalar fields, often referred to in the literature
as ``Q-cloud''.  The mass of the cloud of scalar field, $M_{Q}$,  reads \cite{Herdeiro:2020xmb}~:
\begin{equation}
\label{eq:mq}
M_Q =  \Phi Q + M_{\psi} \ \ , \ \ M_{\psi} = 2 \int\limits_{x_0}^{\infty} {\rm d} x \ x^2 
\sigma \left(\frac{e^2 V^2 \psi^2}{N \sigma^2} -U(\psi)\right).
\end{equation}
where $x_0=0$ for globally regular solutions and $x_0=x_h$ for black holes. The first and second terms represent the contributions
of  the electromagnetic and scalar fields respectively.

Finally, black holes possess properties equivalent to that of thermodynamic systems and we can define the analogue of a temperature $T_H$
and an entropy $S$, respectively, for the solutions that possess a horizon. For the model discussed here the explicit expressions are
\begin{equation}
T_H=\frac{1}{4\pi} \sigma(x_h) \left. N'\right\vert_{x=x_h} \ \ , \ \ S= \frac{1}{4} A_H = \pi x_h^2 \ . 
\end{equation}
Both globally regular as well as black hole solutions obey a Smarr law which reads \cite{Herdeiro:2020xmb}~:
\be
\label{smarr}
     M = \alpha( \Phi Q + M_{\psi} ) \ \ \ , \ \ \  M = \frac{1}{2} T_H A_H + \alpha( \Phi Q + M_{\psi} )
\ee
where $M$ is the ADM mass.

Let us also note that the null energy condition for the solutions studied here reads
$-T_t^t + T_{i}^i  \geq 0$, $i=1,2,3$, i.e. becomes
\begin{equation}
N \psi'^2 + \frac{(\omega - V)^2 \psi^2}{N\sigma^2} \geq 0 \ \ , \ \ 
\frac{V'^2}{2\sigma^2} + \frac{(\omega - V)^2 \psi^2}{N\sigma^2} \geq 0  \ .
\end{equation}
Obviously, the conditions are fulfilled for all solutions that we present in the following.

In \cite{Brihaye:2021phs}, we have shown that the black hole solutions of this model can possess what we called ``wavy'' scalar hair,
i.e. solutions that show a spatial oscillation of the scalar field well outside the horizon of the black hole.
The oscillations also appear in the other fields. In the following, we will demonstrate that this behaviour also
appears for boson stars, i.e. strongly gravitating, but globally regular solutions and does not rely on the existence of a horizon in the space-time.

\section{Boson stars with wavy scalar hair}

It is straightforward to show that the equations (\ref{eq:einstein}), (\ref{eq:matter}) allow for globally regular solutions, i.e. solutions
that exist on the interval $x\in [0:\infty)$. The behaviour at $x\rightarrow \infty$ is equivalent to that for
black holes and determined by the global charges of the solution. The behaviour at $x \ll 1$, on the other hand, reads~:
\begin{equation}
\label{eq:expansion_matter_x_small}
V(x) = V_0 + \frac{e^2 V_0^2 \psi_0^2}{3} x^2 + O(x^3) \ \ , \ \ 
\psi(x) = \psi_0 + \frac{1}{6}
\left. \left(\frac{{\rm d}U}{{\rm d}\psi}\right\vert_{\psi=\psi_0}   - \frac{e^2 \psi_0 V_0^2}{\sigma_0^2}\right) x^2 + O(x^3)
\end{equation}
for the matter fields and 
\begin{equation}
\label{eq:expansion_metric_x_small}
 m(x) = \frac{\alpha}{3} \left(U(\psi_0) + \frac{(\psi_0 V_0)^2}{\sigma_0^2}\right) x^3 + O(x^4) \ \ , \ \
\sigma(x) = \sigma_0\left( 1 + \frac{\alpha e^2 \psi_0^2 V_0^2}{\sigma_0^2}x^2\right) + O(x^3) \ .
\end{equation}
for the metric functions.  The  expansion is given in terms of the parameters $V_0$, $\psi_0$, $\sigma_0$, which have to be determined numerically. These expansions then suggest the following boundary conditions in order to find a globally regular solution to the equations (\ref{eq:einstein}), (\ref{eq:matter}):
\begin{equation}
\label{eq:bc_matter}
    \psi(0) = \psi_0 \ \ , \ \   \psi'(0) = 0 \ \ , \ \ V'(0) = 0 \ \ , \ \ \psi(\infty) = 0 \ \  
\end{equation}
for the scalar and electromagnetic field functions and
\begin{equation}
\label{eq:bc_metric}
m(0)=0 \ \ , \ \  \sigma(\infty)=1
\end{equation}
for the metric functions. 
Note that the condition $\psi(0) = \psi_0$ can be replaced by a condition that fixes the electric charge of the solution and reads
\begin{equation}
\lim\limits_{x\to\infty} \left(x^2 V'(x)\right)=Q \ .
\end{equation}

\begin{figure}[ht!]
\begin{center}
{\includegraphics[width=5cm,angle=-90]{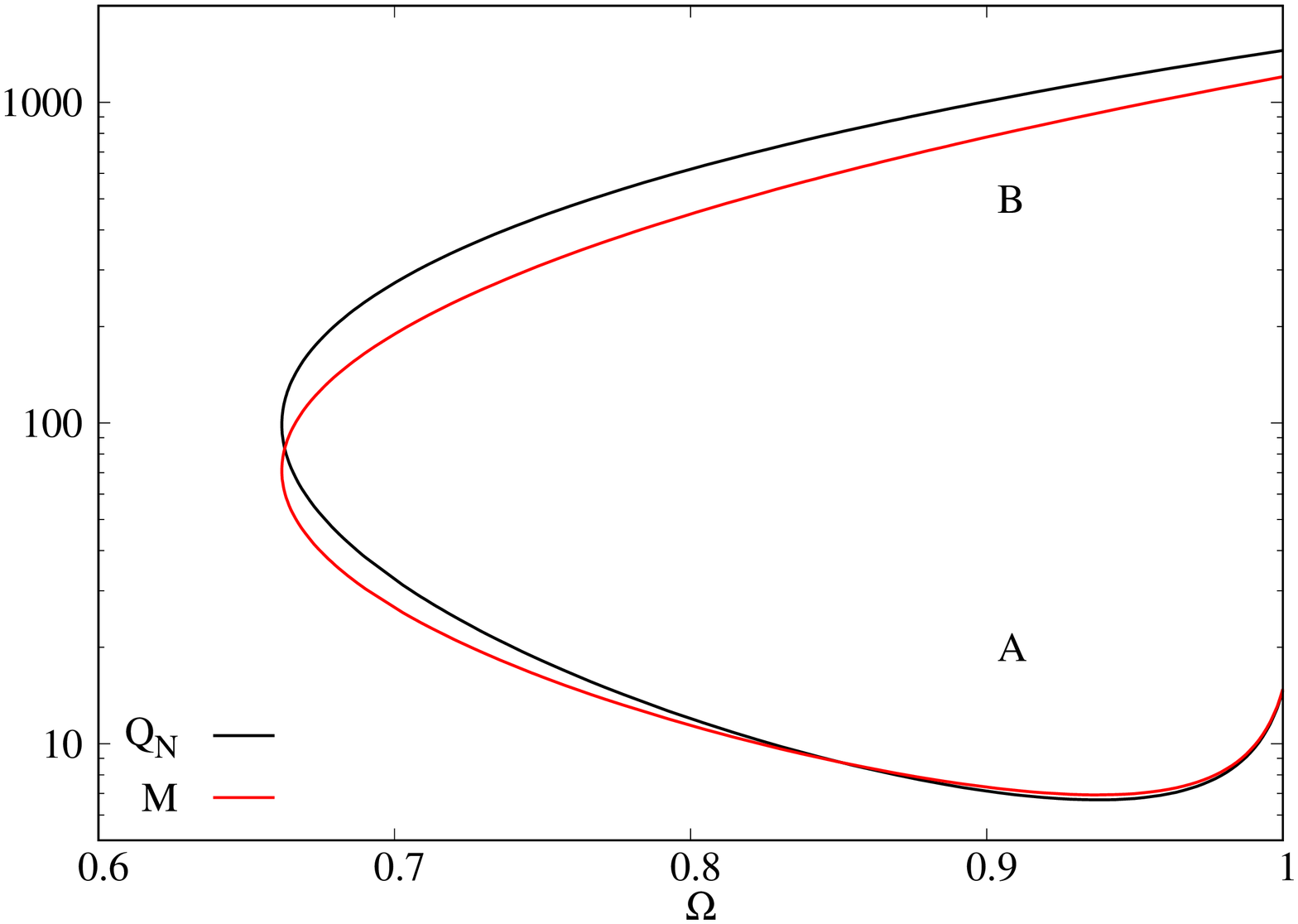}}
{\includegraphics[width=5cm, angle=-90]{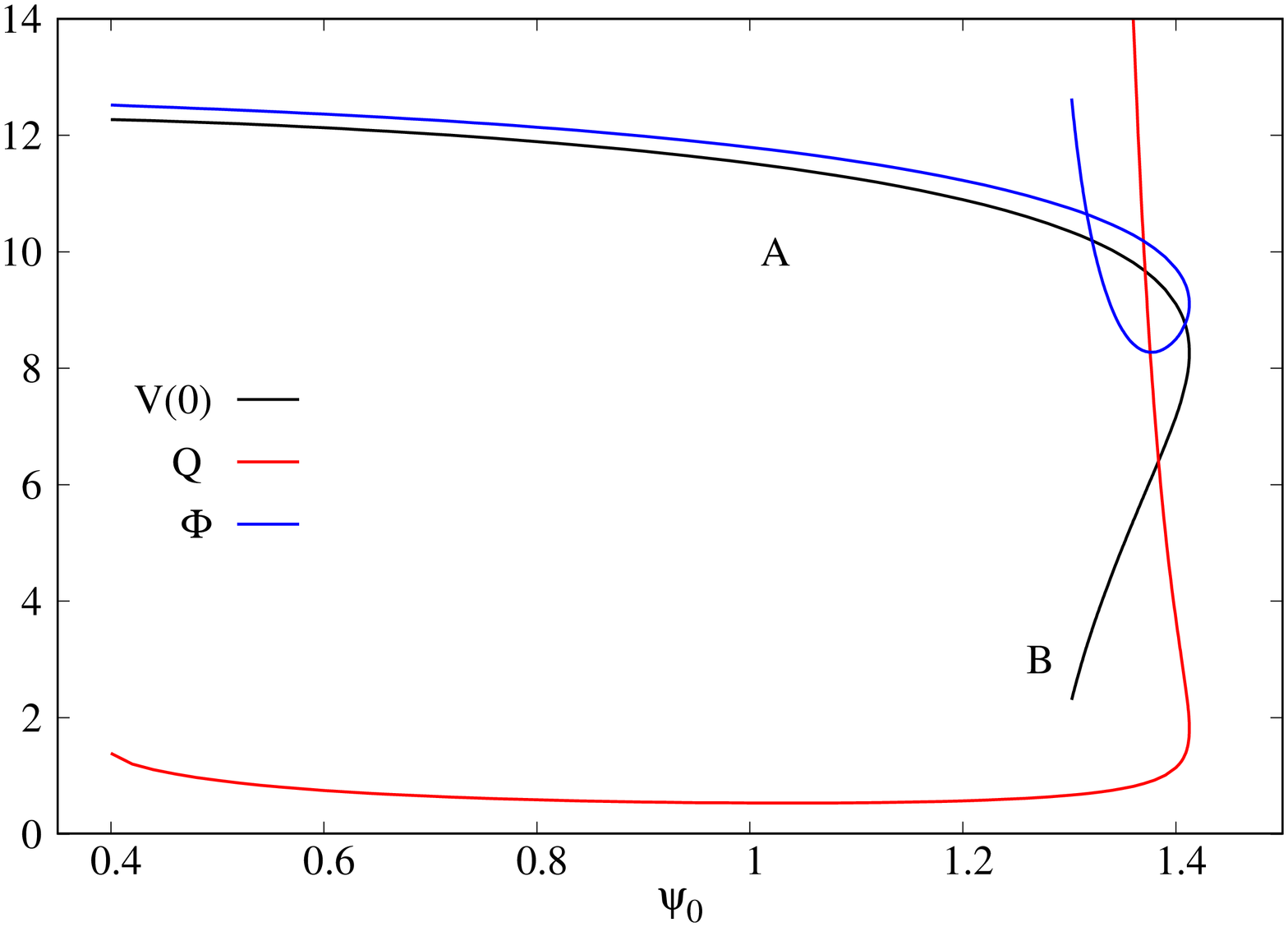}}
\caption{{\it Left}: The  dependence of the mass $M$ and Noether charge $Q_N$ on $\Omega$ 
for Q-balls with
 $e = 0.08$ (and $\beta = 9/32$).
{\it Right}: 
The dependence of the electric charge $Q$, the value $V(0)\equiv V_0$ and  
$\Phi$ on $\psi_0$ for the same set of solutions. 
\label{fig_1}
}
\end{center}
\end{figure}

Let us remind the reader in the following of the basic properties of the solutions in both flat and curved space-time, respectively.
Setting $\alpha=0$, the solutions correspond to charged Q-balls 
and these exist on a finite interval of the parameter $\psi_0$,
i.e. $\psi_0 \in [\psi_{0,1},\psi_{0,2}]$. On the boundaries of this interval, we find
$\Omega \equiv e \Phi \to 1$. This is demonstrated in Fig.\ref{fig_1}, where we give the values of the Noether charge $Q_N$ and the ADM mass $M$ in function of $\Omega$ (left) and the electric charge $Q$, the value of $v(0)$ as well as the asymptotic value of the electric field, $\Phi$, 
in dependence on $\psi_0$ (right) for $e=0.08$. As is obvious from these plots,
we find two branches of solutions in $\Omega$ (in what follows labelled {\it branch A} and {\it branch B}, respectively) which both end at $\Omega =1$ and join at a minimal value of $\Omega$. The solutions on these branches have values of the physical
properties that are very different, e.g. a solution on branch B has much larger mass and Noether charge in comparison to a solution for the same value of $\Omega$ on branch A. 

\begin{figure}[ht!]
\begin{center}
{\includegraphics[width=5cm, angle=-90]{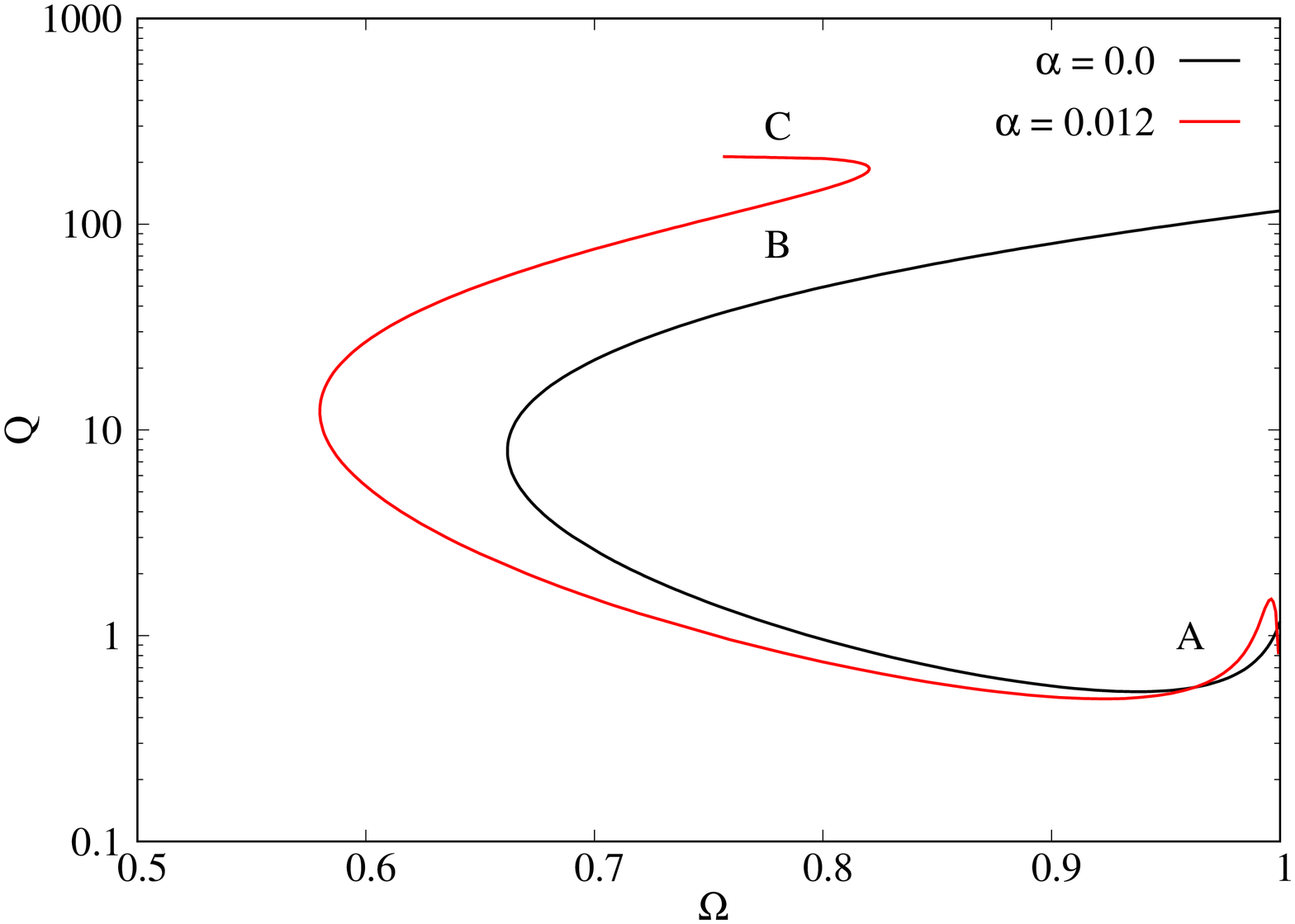}}
{\includegraphics[width=5cm,angle=-90]{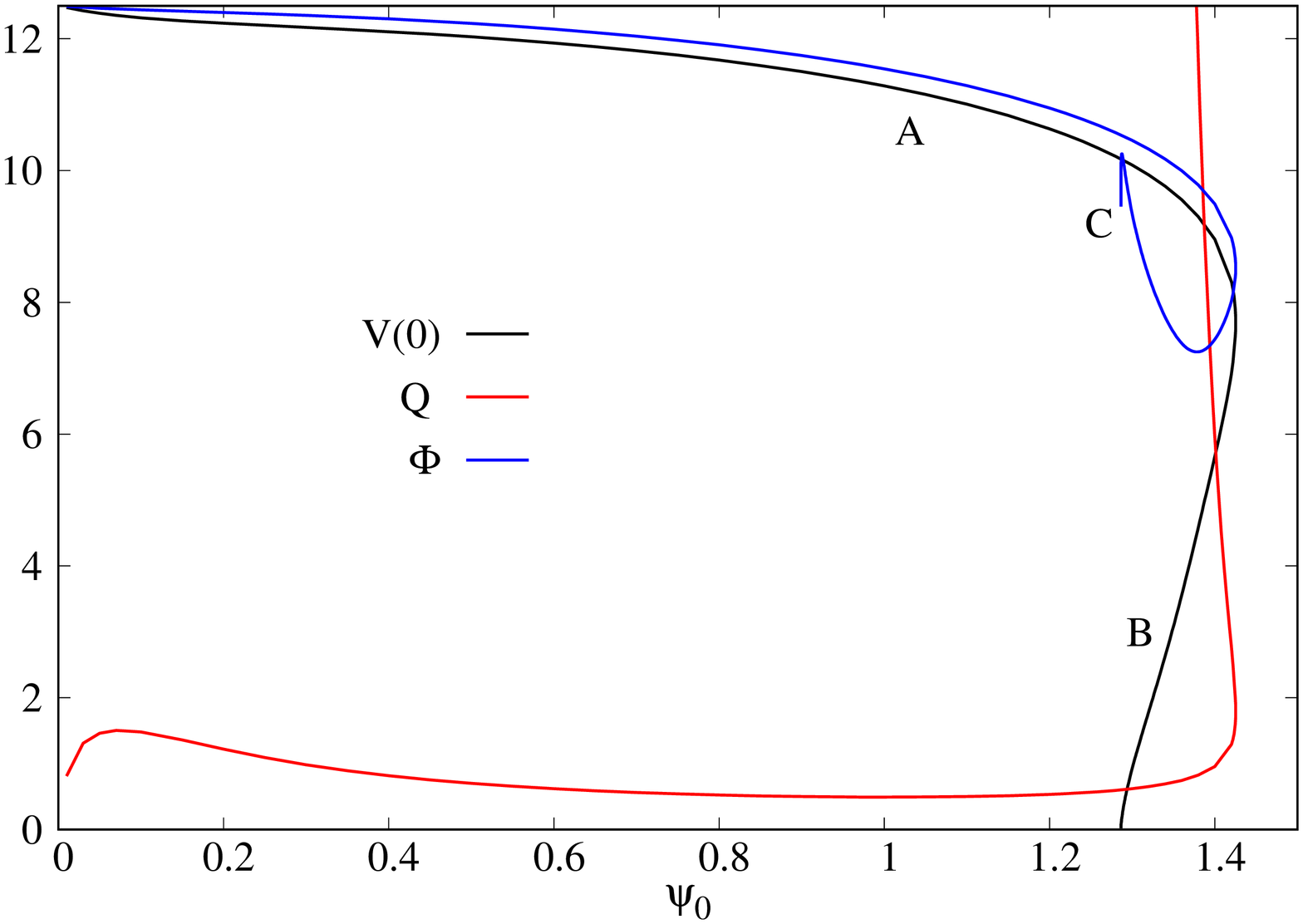}}
\caption{{\it Left}: The dependence of the electric charge $Q$ on the parameter $\Omega\equiv e\Phi$  for $\alpha=0.012$, $e = 0.08$, $\beta = 9/32$.
For comparison, we also give the corresponding curve for $\alpha=0$.
{\it Right}: The dependence of the electric charge $Q$, the value $V(0)\equiv V_0$ and the asymptotic value of the electric potential $\Phi$ 
on $\psi_0$ for the same solutions.
\label{fig_2}
}
\end{center}
\end{figure}

When the space-time is assumed dynamical, i.e. $\alpha > 0$, this two-branch pattern of solutions is modified qualitatively. This is shown in Fig. \ref{fig_2}. When plotting the electric charge $Q$ in function of $\Omega\equiv e\Phi$ (left), we find that branch B does not extend all the way back to $\Omega=1$, but stops at an $\Omega < 1$, where it joins a third branch of solutions, which we will refer to in the following as {\it branch C}. When plotting $Q$, $V(0)$ and $\Phi$ as function of $\psi_0$ the new branch seems a natural extension of the branch B - with even larger values of $Q$ and smaller values of $\Phi$, respectively, however a new local maximum for $V(0)$. 

\begin{figure}[ht!]
\begin{center}
{\includegraphics[width=5cm, angle=-90]{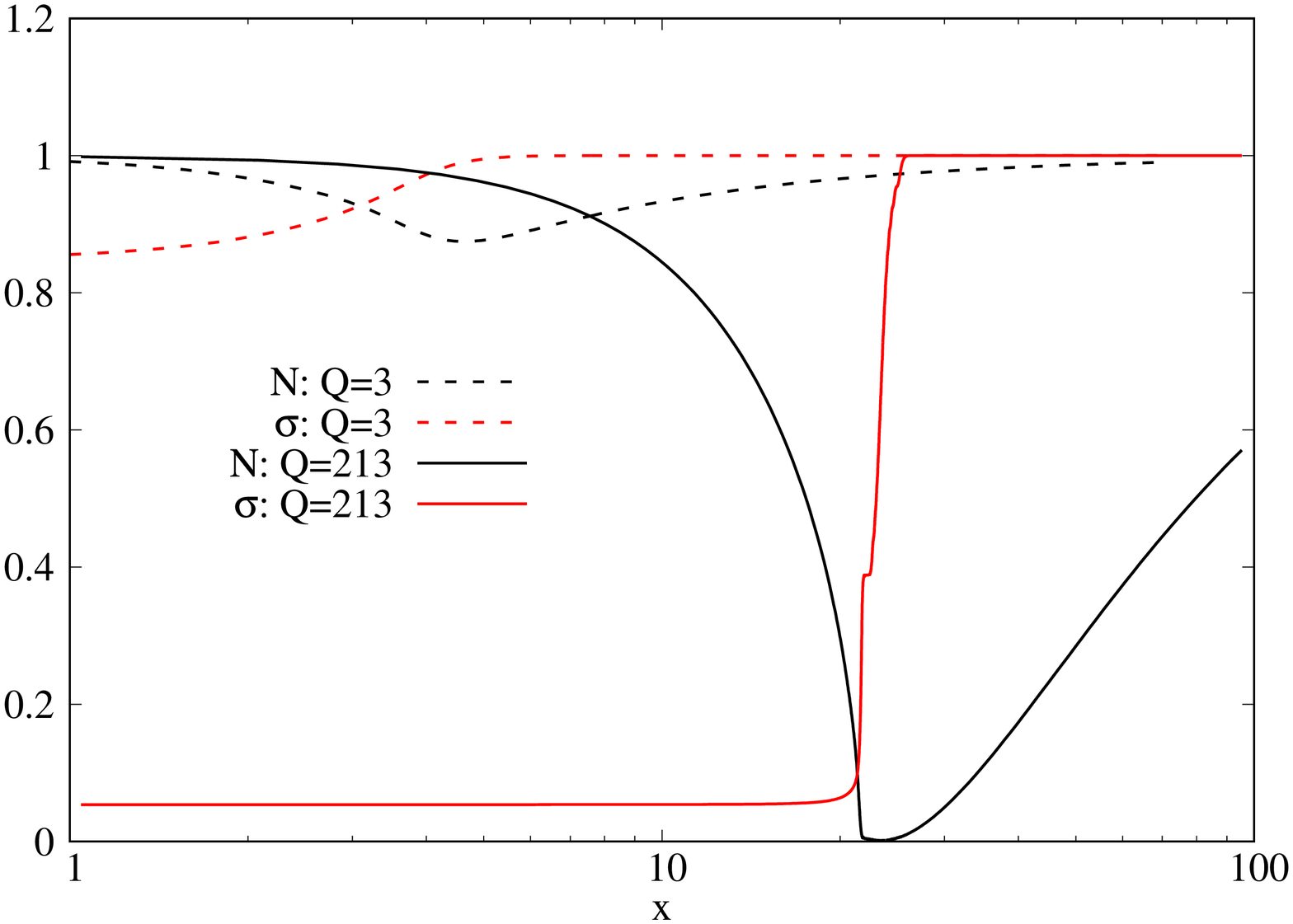}}
{\includegraphics[width=5cm,angle=-90]{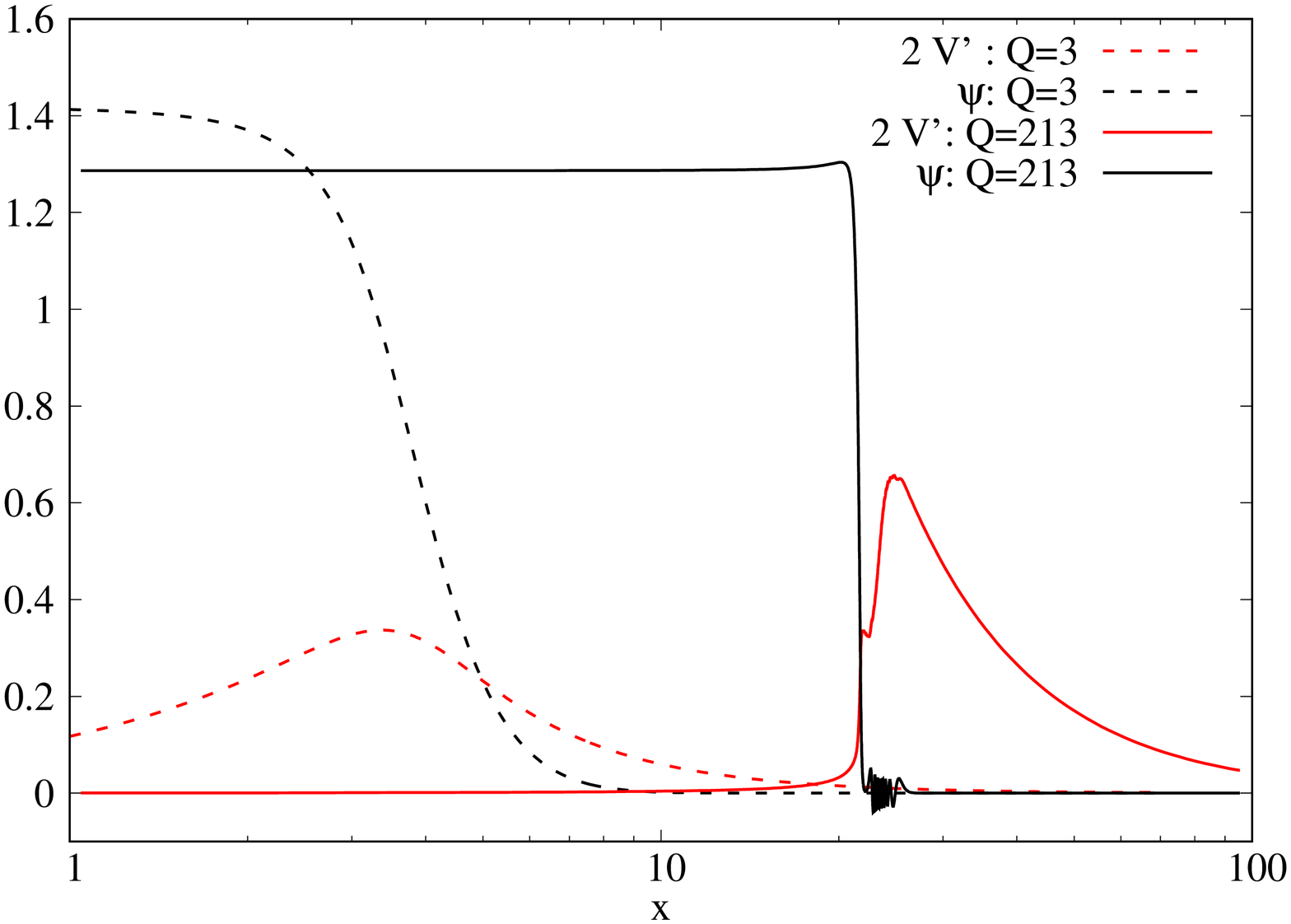}}
\caption{{\it Left}: The profile of the metric functions $N$ and $\sigma$ for a solution on branch A (with $Q=3$) and for a solution on branch C (with $Q=213$), respectively.
For both solutions, we have chosen $\alpha=0.012, e=0.08$.
{\it Right}: The profiles of the corresponding electric field function $V'$ and scalar field function $\psi$.
\label{fig:profile}
}
\end{center}
\end{figure}

In order to understand what distinguishes a solution on branch A from a solution on branch C, we show the profiles  of the metric functions $N$ and $\sigma$ (left) as well as for the profiles of the scalar field function $\psi$ and the electric field function $V$ (right) for a typical solution on branch $A$ (with $Q=3$) and compare it to a typical solution on branch $C$, see Fig.\ref{fig:profile}. Both solutions
are for $\alpha=0.012$, $e=0.08$, but have very different values of the
electric  charge: the solution on branch $A$ has $Q=3$, while the solution on branch $C$ has  $Q=213$. 

Moving from branch A to branch B and finally to branch C corresponds to
an increase in the electric charge $Q$
(or equivalently in the decrease of 
the parameter $V(0)$). Moving from branch A to C, we observe that the minimum of the metric function $N$ deepens and moves away from the center of the boson star. Moreover, the metric function $\sigma$ at the center of the star decreases and develops a step-like profile from an originally approximately linear increasing function in the transition region between the interior of the boson star and the asymptotically flat exterior. Following this process
it is found that the metric function $N(x)$ presents a local minimum that decreases in value. 
At the same time the scalar function $\psi(x)$ associated with the branches $A$ and $B$  decreases
monotonically to zero. We observe this step-like behaviour also in the matter field functions. In fact, we find that for the solution on branch C, we can define an interior region $x\in [0:\tilde{x}]$ for which
$\psi(x)\equiv p_0\neq 0$, where $p_0$ is a positive constant. In this interval for $x$ we also have that the electric field $V'(x)\equiv 0$. 
This is very different from the solution on branch A, where the scalar field decays smoothly from its central value to zero with a nearly
linear decay in the intermediate region, where the electric field possesses a local maximum. In fact, as pointed out already in 
\cite{Brihaye:2020vce}, the constant scalar field generates
constant potential scalar field energy density that can be interpreted as a positive cosmological constant.

\begin{figure}[ht!]
\begin{center}
{\includegraphics[width=5cm, angle=-90]{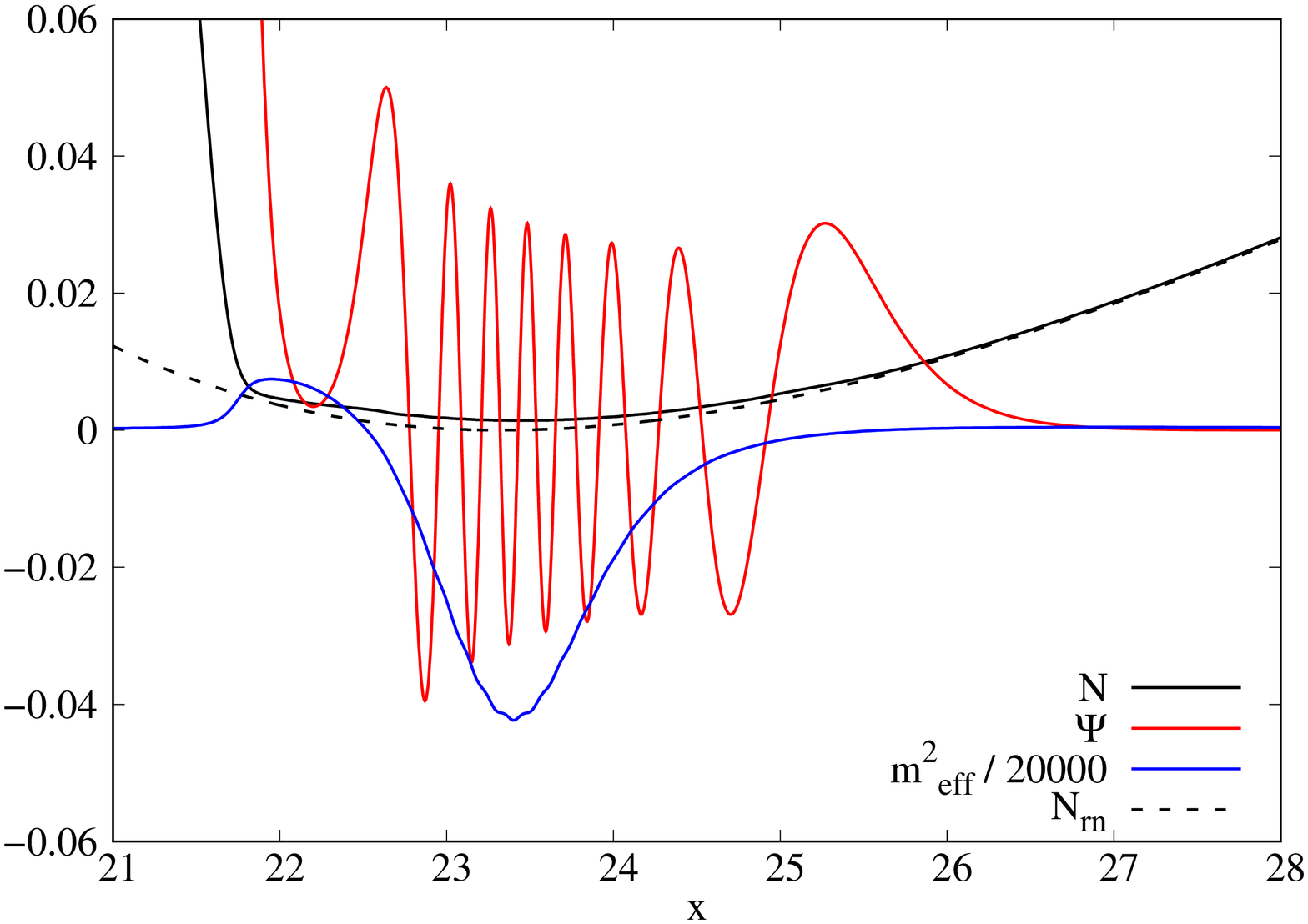}}
{\includegraphics[width=5cm,angle=-90]{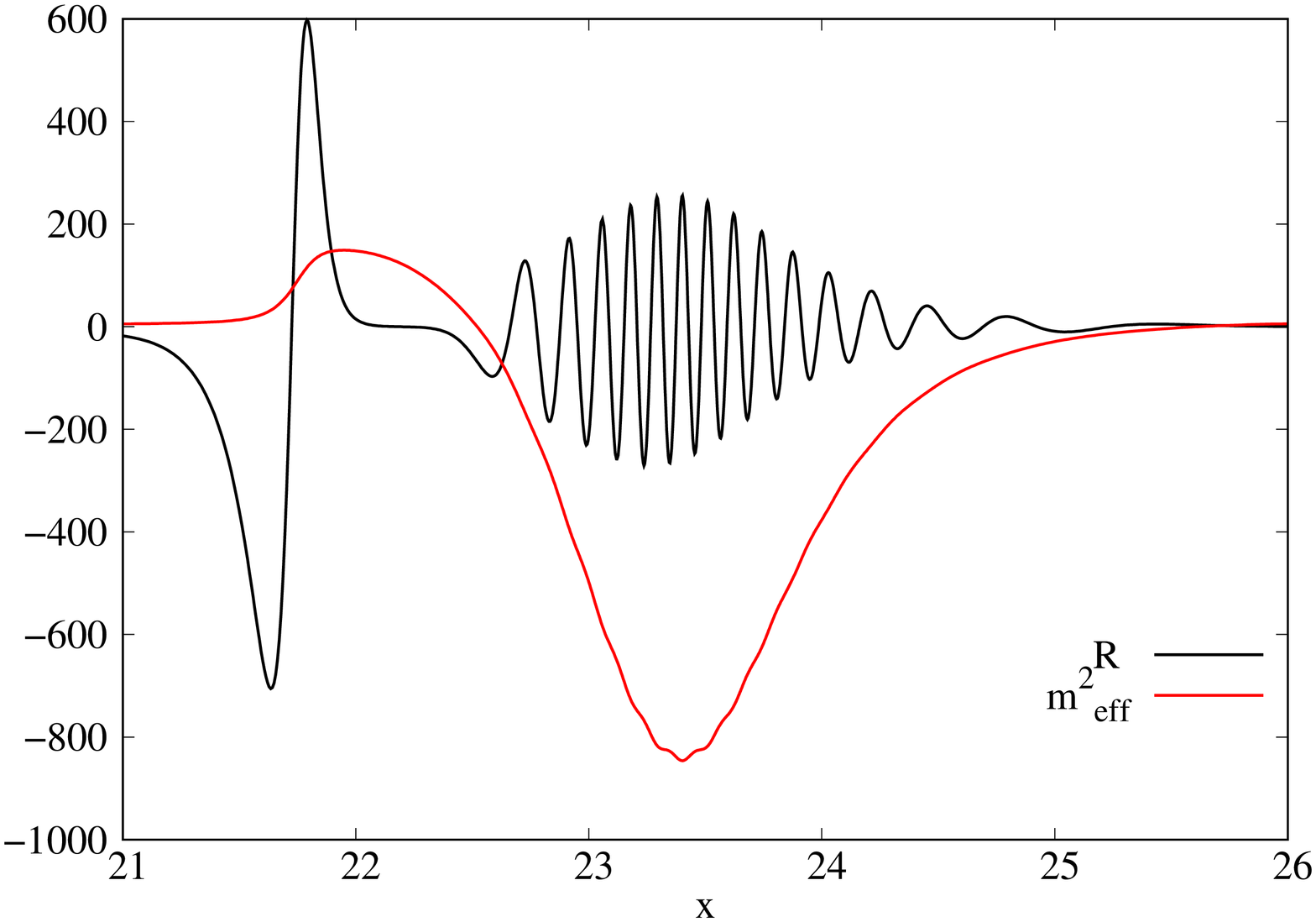}}
\caption{{\it Left}: Details  of the functions $N$ and $\psi$ for the solution on branch C shown in Fig. \ref{fig:profile}
in the region of the local minimum of $N(x)$. For comparison, we also
show the metric function $N(x)$ of the extremal Reissner-Nordstr\"om solution
with the same mass. We also give the effective mass $(m_{\rm eff})^2$.
{\it Right}: The corresponding Ricci scalar $R$ as well as
the effective mass $(m_{\rm eff})^2$. \label{fig:profile_zoom}
}
\end{center}
\end{figure}

An interesting new feature appears for $x  > \tilde{x}$. For $x > \bar{x}$, the scalar field function $\psi(x)\equiv 0$ such that the space-time can be described by a Reissner-Nordstr\"om space-time for $x > \bar{x}$. However, for specific combinations of $\alpha$ and $Q$
we find that there is an intermediate region $x\in [\tilde{x}:\bar{x}]$
in which the scalar field shows spatial oscillations. In fact, we have discussed this phenomenon for the first time in \cite{Brihaye:2021phs} for black holes and demonstrate here that it exists also for boson stars, see Fig. \ref{fig:profile_zoom}.

\begin{figure}[ht!]
\begin{center}
{\includegraphics[width=5cm, angle=-90]{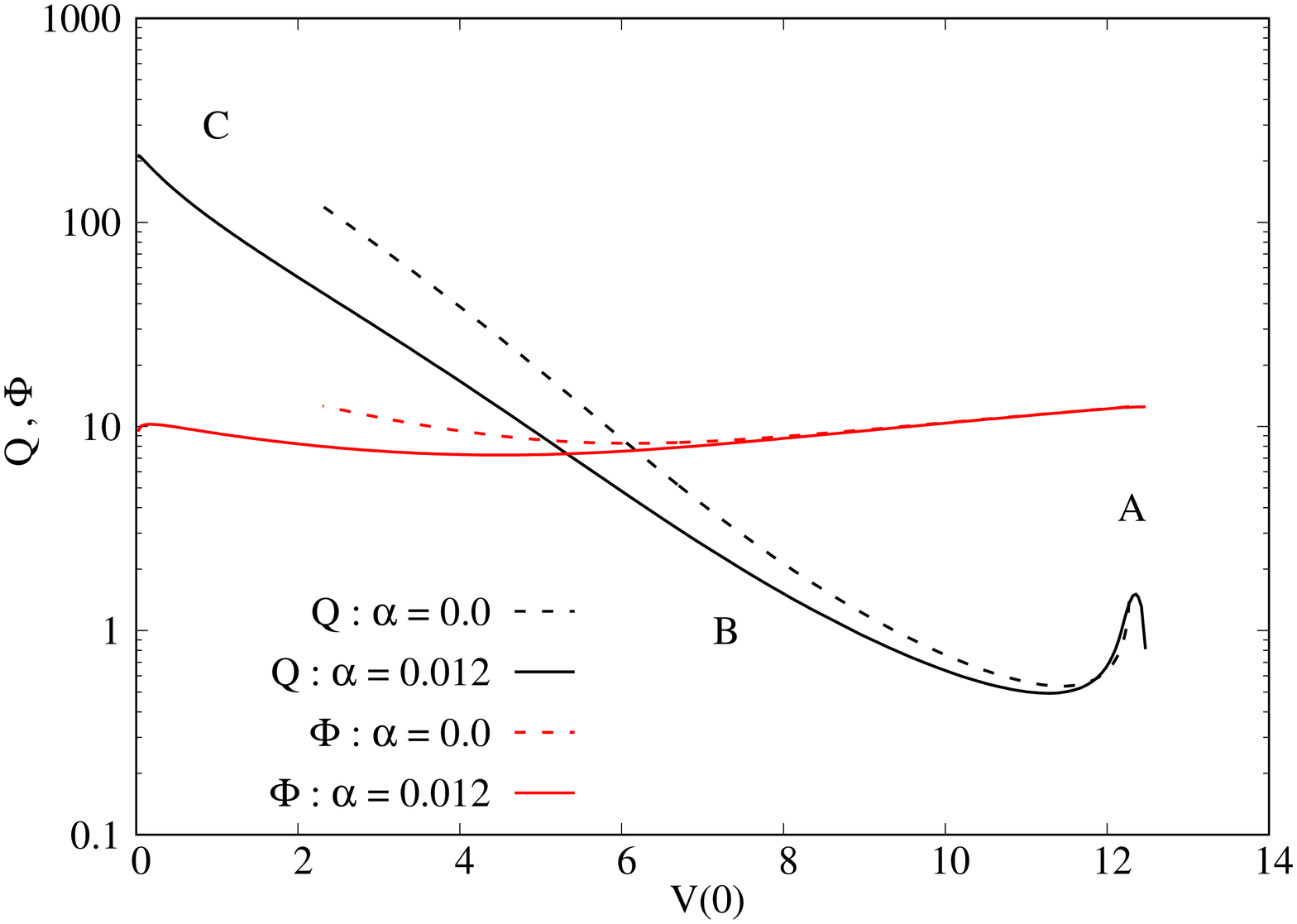}}
{\includegraphics[width=5cm,angle=-90]{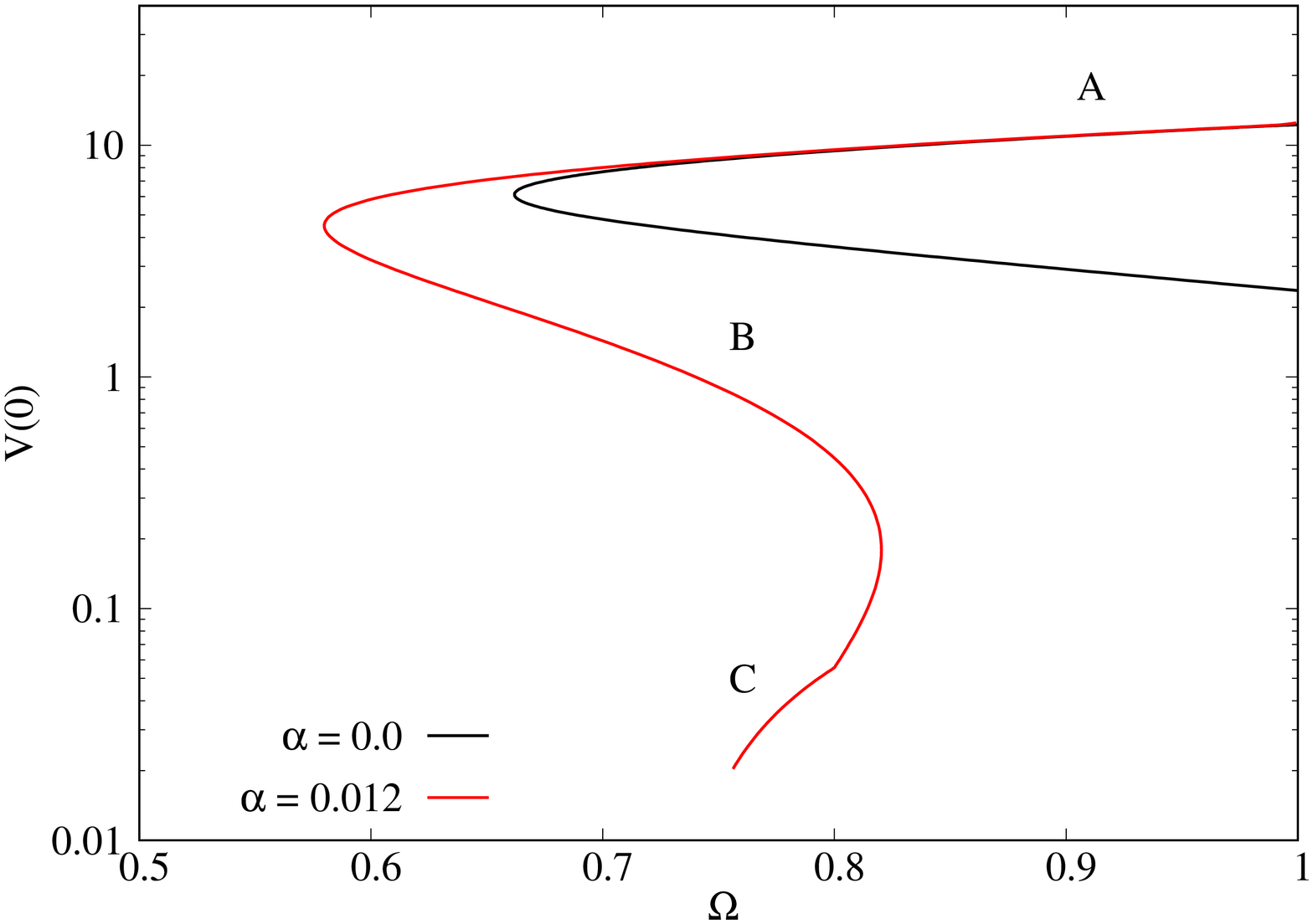}}
\caption{{\it Left}: The dependence of the charge $Q$ and the asymptotic value of the electric potential $\Phi$
on the parameter $V(0)$  for the same set of solutions as shown in Fig.\ref{fig_2}. 
{\it Right}: The  dependence of $V(0)$ on $\Omega$ for the same set of solutions.
\label{fig_3}
}
\end{center}
\end{figure}

In fact, one can understand this behaviour by considering the scalar field equation and assuming for simplicity that 
the oscillations of the scalar field are small enough to neglect the terms in the scalar field equation that are higher than linear order. The equation then reads
(note that this is valid only for $x\in[\tilde{x}:\bar{x}]$)
\be
\label{eq:effective_mass}
         \frac{1}{x^2 \sigma N} (x^2 \sigma N \psi')' =  m_{eff}^2(x) \psi \ \ , \ \ 
				 m_{eff}^2(x) = \frac{1}{N(x)} - \frac{e^2 V^2(x)}{N^2(x) \sigma^2(x)} \ \ , \ 
\ee
where $m_{eff}^2$ now plays the role of a position-dependent mass of the scalar field $\psi$. Our numerical results show that $N(x)$ is very small, but non-zero for $x\in[\tilde{x}:\bar{x}]$, for the solution on branch C discussed above we find typically $N(x) \sim 0.005$ for $x \in [22,25]$. Hence $m_{eff}^2$ has large negative values and (\ref{eq:effective_mass}) allows for oscillating solutions.
All these features are illustrated in Fig. \ref{fig:profile_zoom}, where details of the solution of branch $C$ shown in Fig. \ref{fig:profile} are given. The comparison of the metric function $N(x)$ with the corresponding metric function $N_{\rm rn}(x)$ of the extremal Reissner-Nordstr\"om solution with the same mass (left) demonstrates that
$N(x)$ is very close to $N_{\rm rn}(x)$, however, does not possess
a double zero because it stays always positive (albeit small). 
The interesting feature about the oscillations in the scalar field are that they lead to oscillations in the curvature scalars, see Fig.\ref{fig:profile_zoom} (right) for the profile of the Ricci scalar $R$.

\begin{figure}[ht!]
\begin{center}
{\includegraphics[width=5cm, angle=-90]{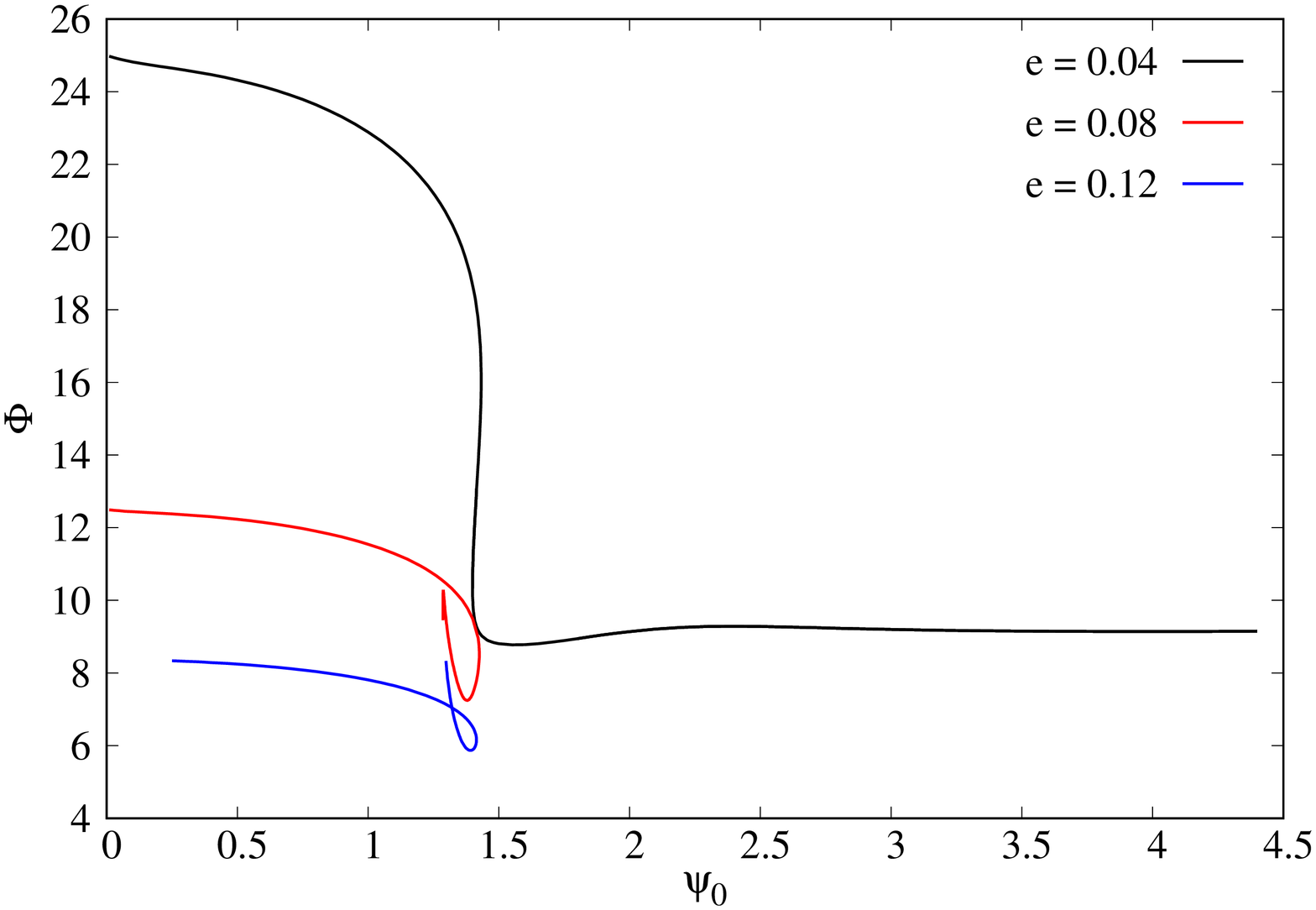}}
{\includegraphics[width=5cm,angle=-90]{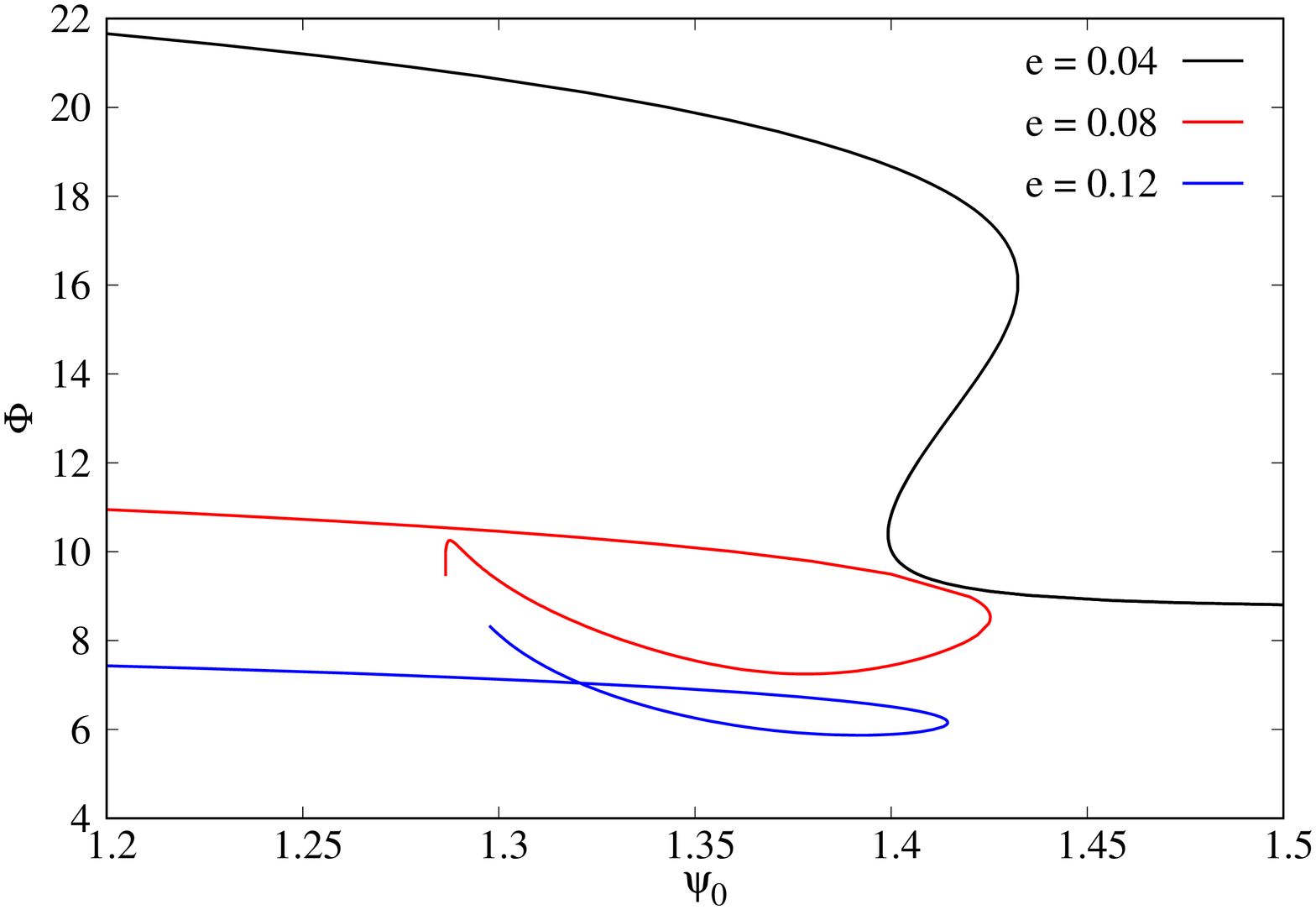}}
\caption{{\it Left}: We show the dependence of the parameter $\Phi$ on $\psi_0$ for three values of $e$
and  $\alpha = 0.012$.
{\it Right}: Details on the data show left  in the region $\psi_0 \in [1.2,1.5]$.
\label{fig_data_phi_0_phi}
}
\end{center}
\end{figure}

\begin{figure}[ht!]
\begin{center}
{\includegraphics[width=5cm, angle=-90]{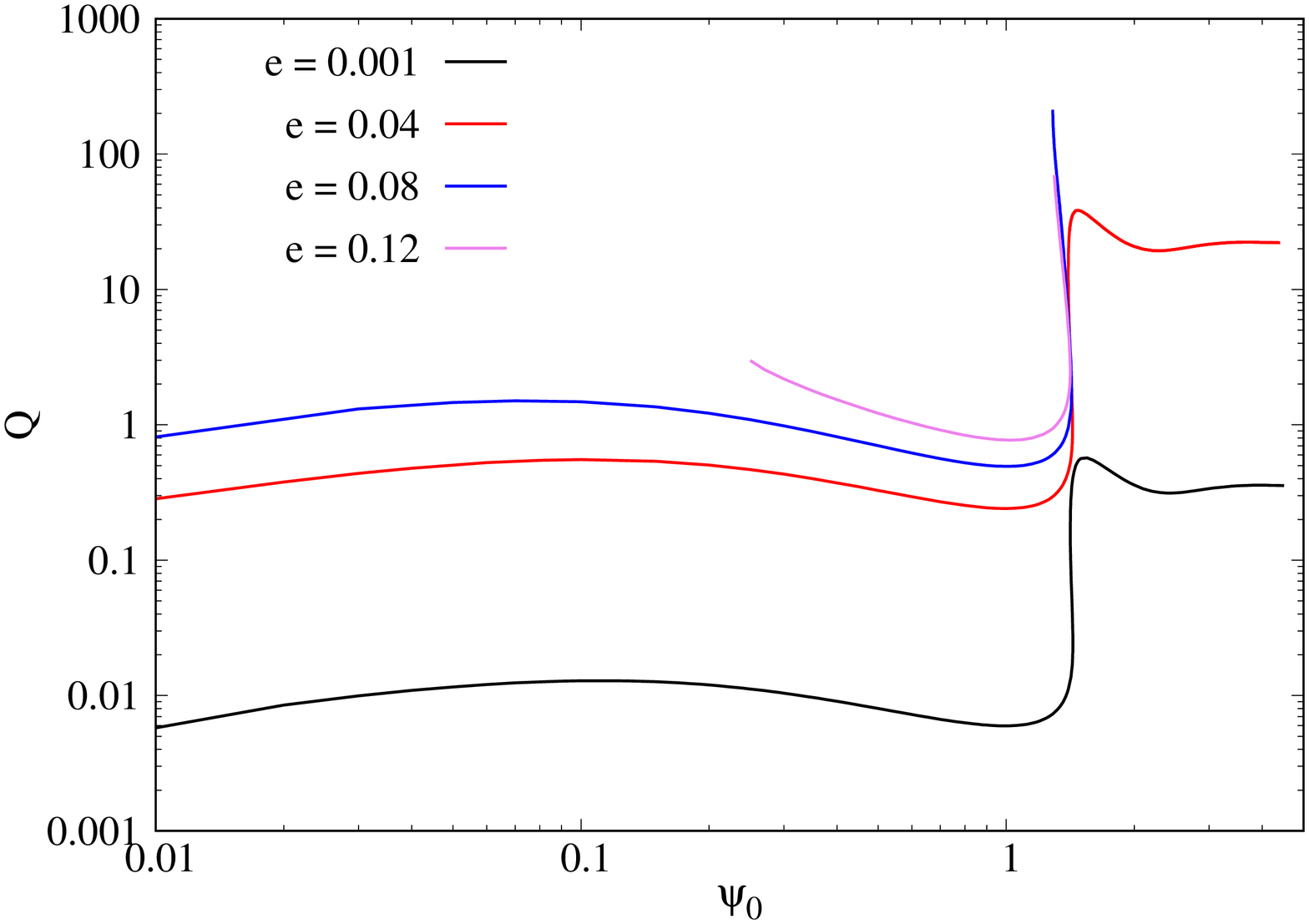}}
{\includegraphics[width=5cm,angle=-90]{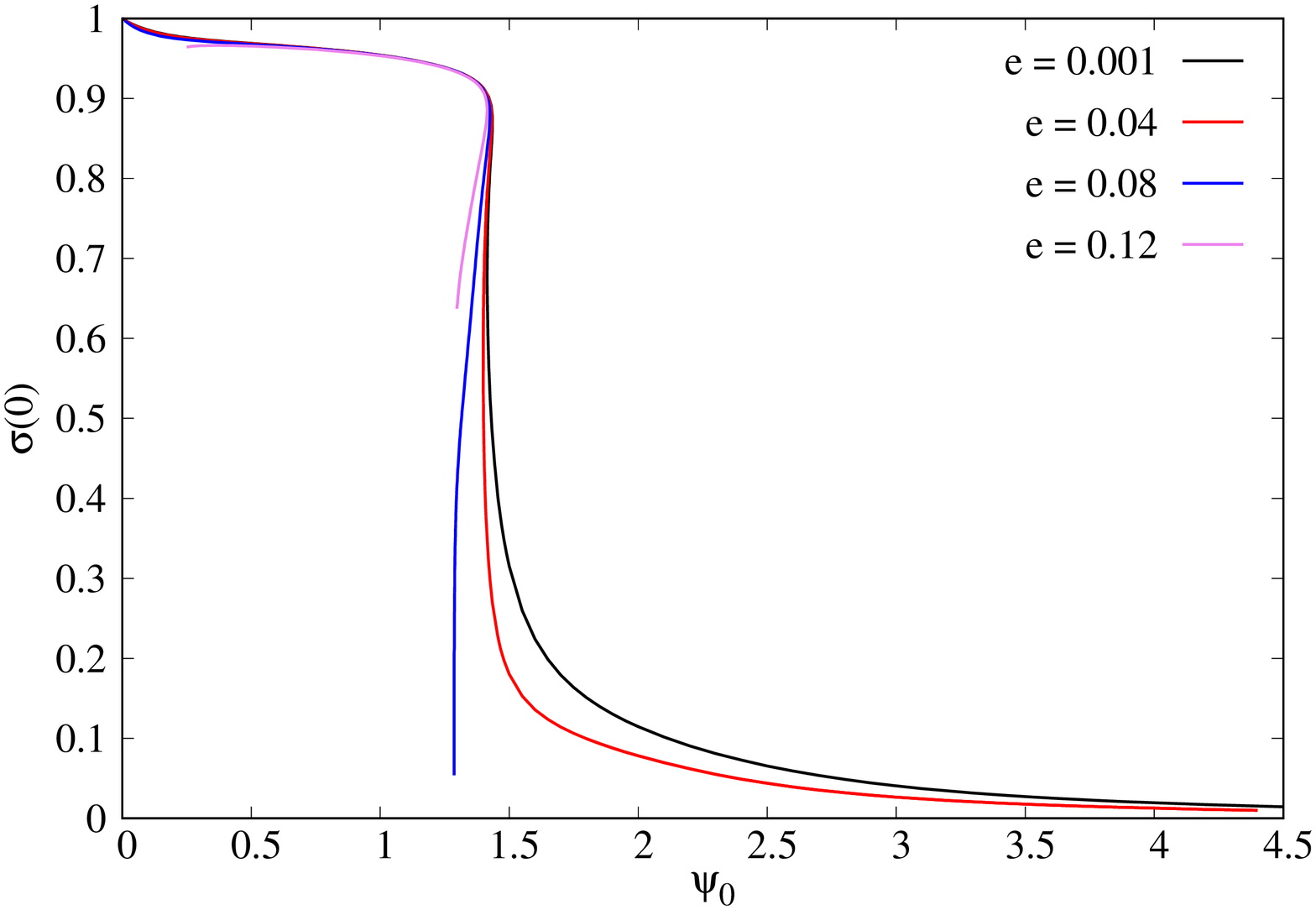}}
\caption{{\it Left}: We show the dependence of the electric charge $Q$ on the parameter $\psi_0$ for three values of $e$
and  $\alpha = 0.012$. {\it Right}: We show the dependence of $\sigma(0)$ on $\psi_0$ for the same set of solutions.
\label{fig_data_phi_0_charge}
}
\end{center}
\end{figure}

\begin{figure}[ht!]
\begin{center}
{\includegraphics[width=5cm, angle=-90]{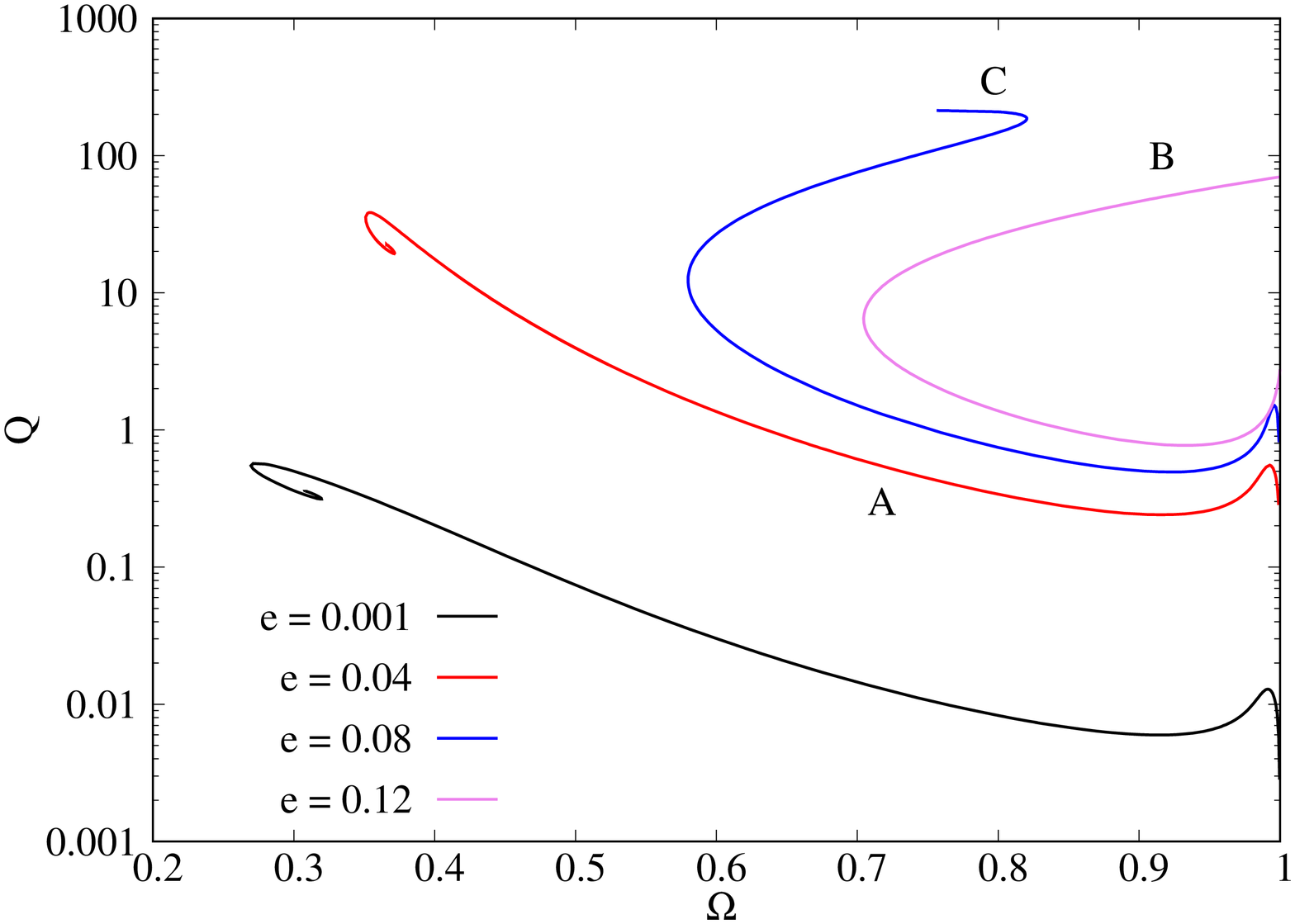}}
{\includegraphics[width=5cm,angle=-90]{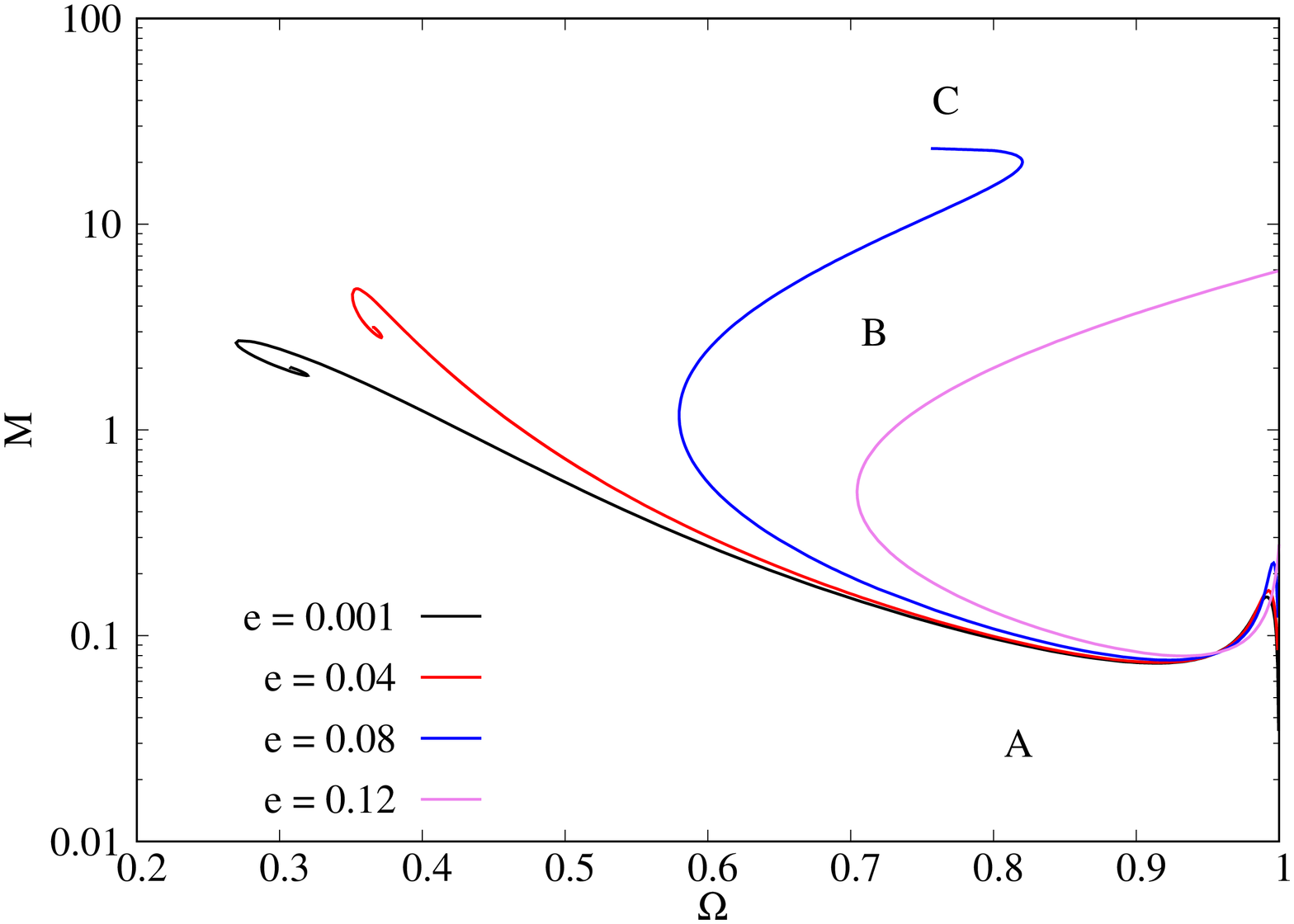}}
\caption{{\it Left}: The dependence of the electric charge $Q$ on $\Omega$ for three values of $e$
and  $\alpha = 0.012$.
{\it Right}: The dependence of the mass $M$ on $\Omega$ for the same set of solutions.
\label{fig_data_ome_q_m}
}
\end{center}
\end{figure}

For completeness, we also demonstrate that while the solutions
on branch C possess large values of the electric charge, they possess small values of $V(0)$ as compared to the corresponding solutions on branch A. This is shown in Fig. \ref{fig_3}, where we give the
dependence of the charge $Q$ and the asymptotic value of the electric potential $\Phi$
on the parameter $V(0)$ as well as the dependence of $V(0)$ on $\Omega$
for the same set of solutions as given in Fig.\ref{fig_2}.\\
\\
For all the solutions discussed above, we have fixed the gauge coupling constant $e=0.08$. In the following, we would like to discuss how the variation of this coupling changes our results. This is shown in 
Figs. \ref{fig_data_phi_0_phi}, \ref{fig_data_phi_0_charge} and \ref{fig_data_ome_q_m} for $\alpha = 0.012$ and several values of $e$.
Our investigation suggests the following: 
\begin{itemize}
\item For small values of $e$ (typically  $e \sim 0.04$) the solutions exist for $\psi_0 \in [0, \psi_{0,{\rm max}}]$
with $\psi_{0,{\rm max}} \approx 4.5$. 
In the limit $\psi_0 \to \psi_{0,{\rm max}}$, the value $\sigma(0)$ tends to zero and the solution develops a singularity at the center.
Note that several solutions corresponding to the same value of $\psi_0$ exist around $\psi_0 \sim 1.4$ and that we have hence found it more convenient to replace the boundary condition $\psi(0) = \psi_0$ by the alternative boundary condition $\lim\limits_{x\to\infty} \left(x^2 V'(x)\right)=Q$ for the numerical construction of these solutions.

\item For large values of $e$ (typically  $e \sim 0.12$) the branch of solutions stops at some finite value of $\psi_0$ because
$\Omega \rightarrow 1$. In this case, two branches of solutions (branch A and branch B) exist in $\Phi$.

\item For the intermediate values of $e$ (typically  $e \sim 0.08$) we observe the phenomenon of oscillations and existence of branch C as discussed above. 
\end{itemize}

\begin{figure}[ht!]
\begin{center}
{\includegraphics[width=8cm,angle=-90]{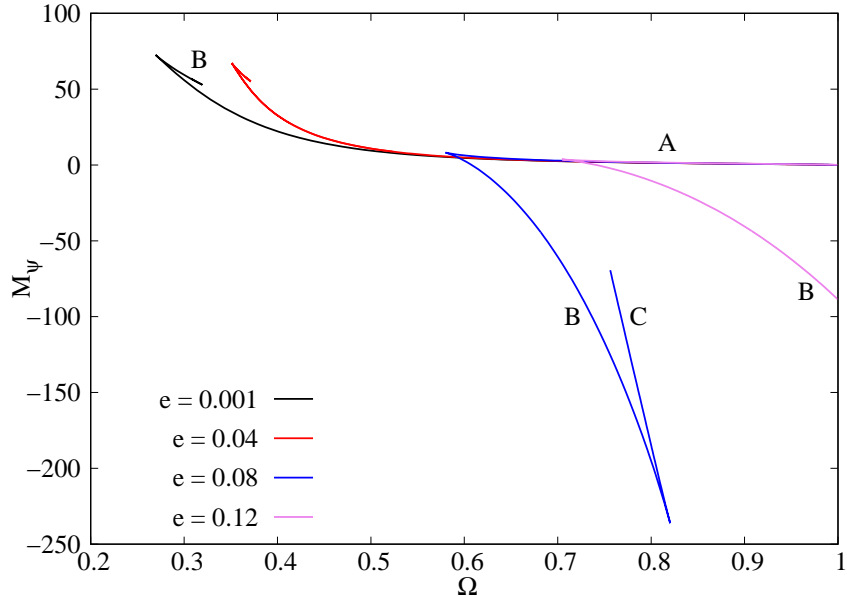}}
\caption{The dependence of the mass $M_{\psi}$ on $\Omega$ for several values of $e$
and  $\alpha = 0.012$.
\label{mass_psi}
}
\end{center}
\end{figure}

Finally, we show the dependence of the mass 
$M_{\psi}$ on $\Omega= e \Phi$ in
Fig. \ref{mass_psi} for fixed $\alpha$ and several values of $e$. This demonstrates that for solutions on 
branch $C$ the mass $M_{\psi}$ is proportional to $\Phi$, a property not seen on any of the other branches.
Remembering that the metric  functions of these solutions are very close to those of an
extremal RN solution outside an intermediate value of the radial coordinate at which the horizon forms, say $\bar{x}$, the mass and charge of the solutions are such that
$M \sim \bar{x}$, $Q \sim \bar{x}/ \sqrt{\alpha}$. We believe that these facts about branch $C$ explain the peculiar dependence of the 
physical quantities  observed, e.g. in Fig. \ref{fig_1}. A similar statement is true for the corresponding black hole solutions (see next section).

\newpage
\pagebreak
\section{Black holes with wavy scalar hair: new results}
Black holes with wavy scalar hair have first been discussed in \cite{Brihaye:2021phs}. Here we extend and refine the results
and also discuss the interior of these black holes.

We now have to impose boundary conditions at the horizon of the black hole $x=x_h > 0$ (instead of at the origin of the coordinate system) and we find that the appropriate choice reads
\be
\label{eq:bc_bh}
     m(x_h) = \frac{x_h}{2} \ \ , \ \ N' \psi'|_{x=x_h} = \frac{1}{2} \left. \frac{d U}{d \psi}\right\vert_{x=x_h} \ \ ,  \ \ V(x_h) = 0 \ \ , 
\ee
The first condition is equivalent to $N(x_h)=0$, the second is a regularity condition on the scalar field at the horizon, while the third
results from the {\it synchronization condition}. 

\begin{figure}[ht!]
\begin{center}
{\includegraphics[width=5cm, angle=-90]{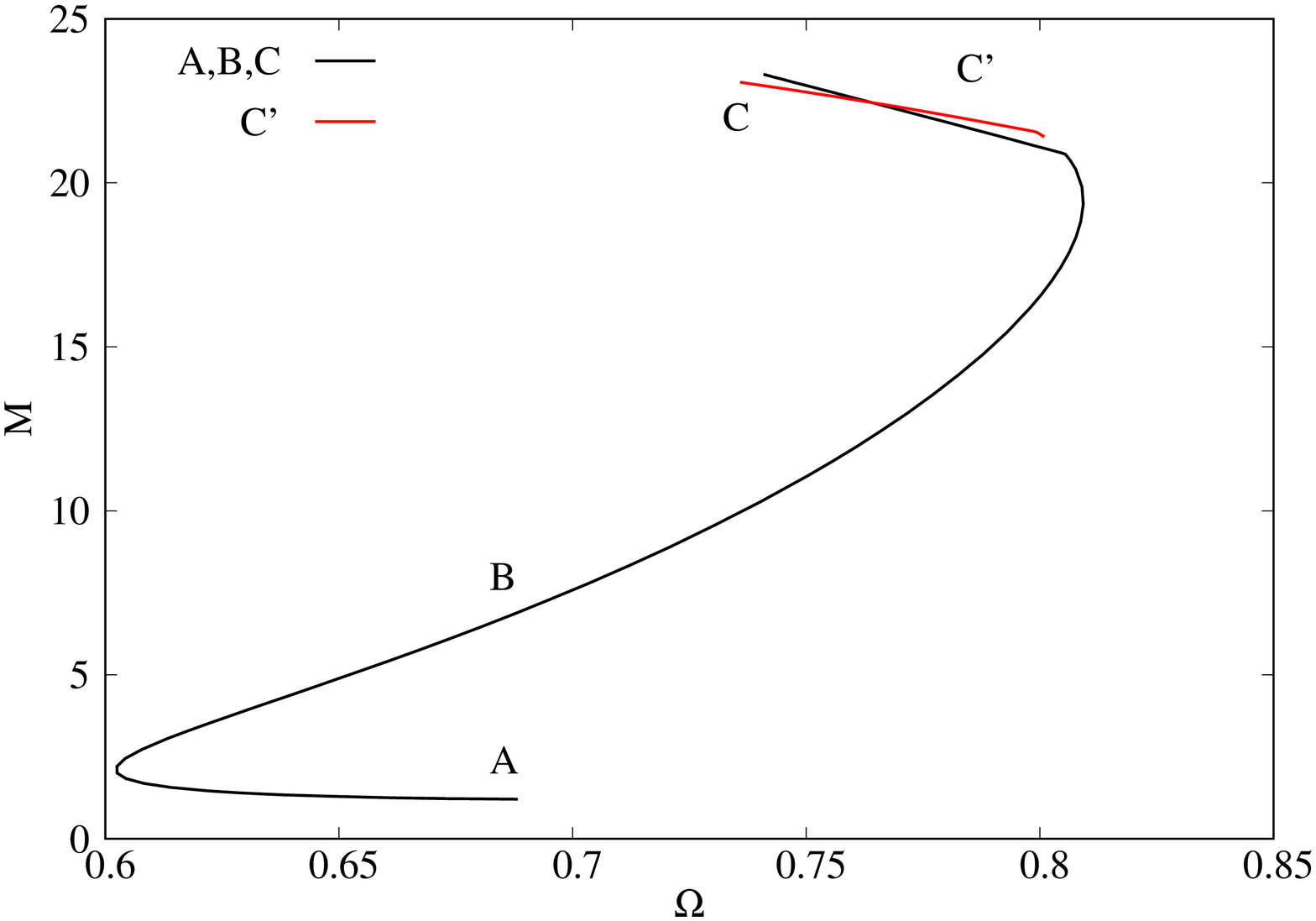}}
{\includegraphics[width=5cm,angle=-90]{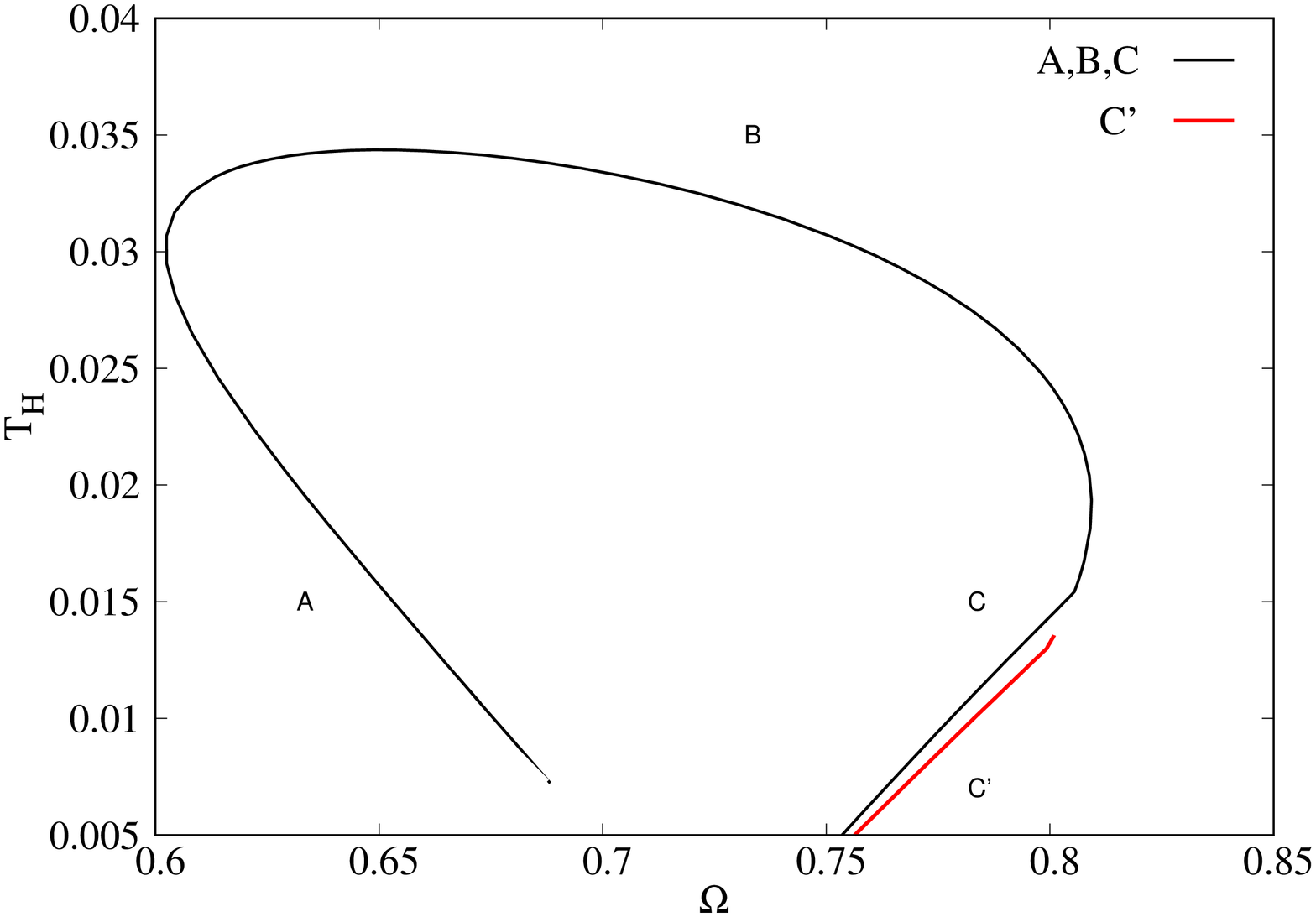}}
\caption{{\it Left}: We show the dependence of the mass $M$ of black holes with scalar hair on $\Omega$ for $\alpha = 0.012$, $e = 0.08$, $x_h=1$. 
{\it Right}: Same as left, but for the temperature $T_H$.
\label{fig:data_rh_1}
}
\end{center}
\end{figure}

As mentioned above, black holes can carry scalar hair in minimally coupled gravity models if a so-called
{\it synchronization condition} is fulfilled. This is a condition imposed on the matter fields at the horizon of the black hole
and in the case we are discussing here reads $\omega-e V(x_h)=0$, where $V(x_h)$ is the value of the gauge field function on the black hole
horizon. Then choosing the gauge $\omega=0$, we have the condition on $V$ given in (\ref{eq:bc_bh}). \\
\\
In the following, we will first discuss the case $\alpha = 0.012$, $e = 0.08$, $x_h=1$. 
We find - in agreement with the results presented for boson stars above - that black holes can form up to three branches
of solutions in $\Omega = e \Phi$. This is shown in Fig. \ref{fig:data_rh_1}, where we show the mass $M$ and the temperature $T_H$
using the same convention for the labeling of the branches. The mass increases along these branches moving from branch A to B and then to C.
On branch A and branch C, respectively, the mass $M$ increases when decreasing $\Omega$, while on branch B the mass increases with increasing
$\Omega$. Comparing this with the dependence of the temperature $T_H$ on $\Omega$, which increases along branch A, but decreases on both branch B and C. We find that this is qualitatively different to the behaviour of charged black holes without scalar hair given by the Reissner-Nordstr\"om solution for which
$\alpha Q^2=x_h(2M -x_h)$ and $4\pi T_H=2(x_h-M)/x_h^2$.
For fixed $x_h$, the charge $Q$ increases with $M$, while the temperature $T_H$ decreases when the mass $M$ is increased. Both branch B and branch C show this qualitative behaviour, while the solutions on branch A show increase in temperature $T_H$ for increasing mass $M$. 
In fact, on branch C, the solutions approach a solution with $T_H=0$.

\begin{figure}[ht!]
\begin{center}
{\includegraphics[width=5cm, angle=-90]{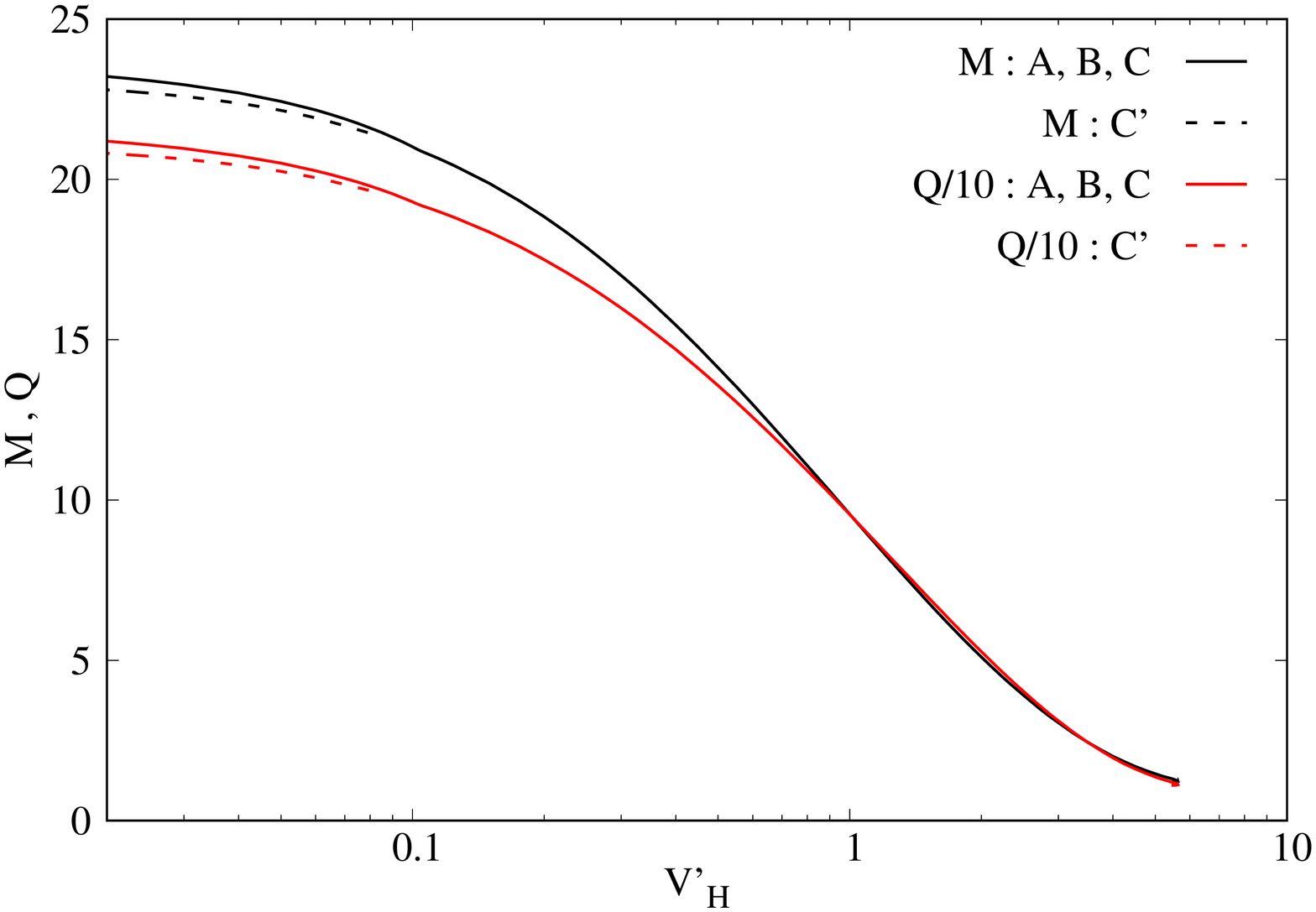}}
{\includegraphics[width=5cm,angle=-90]{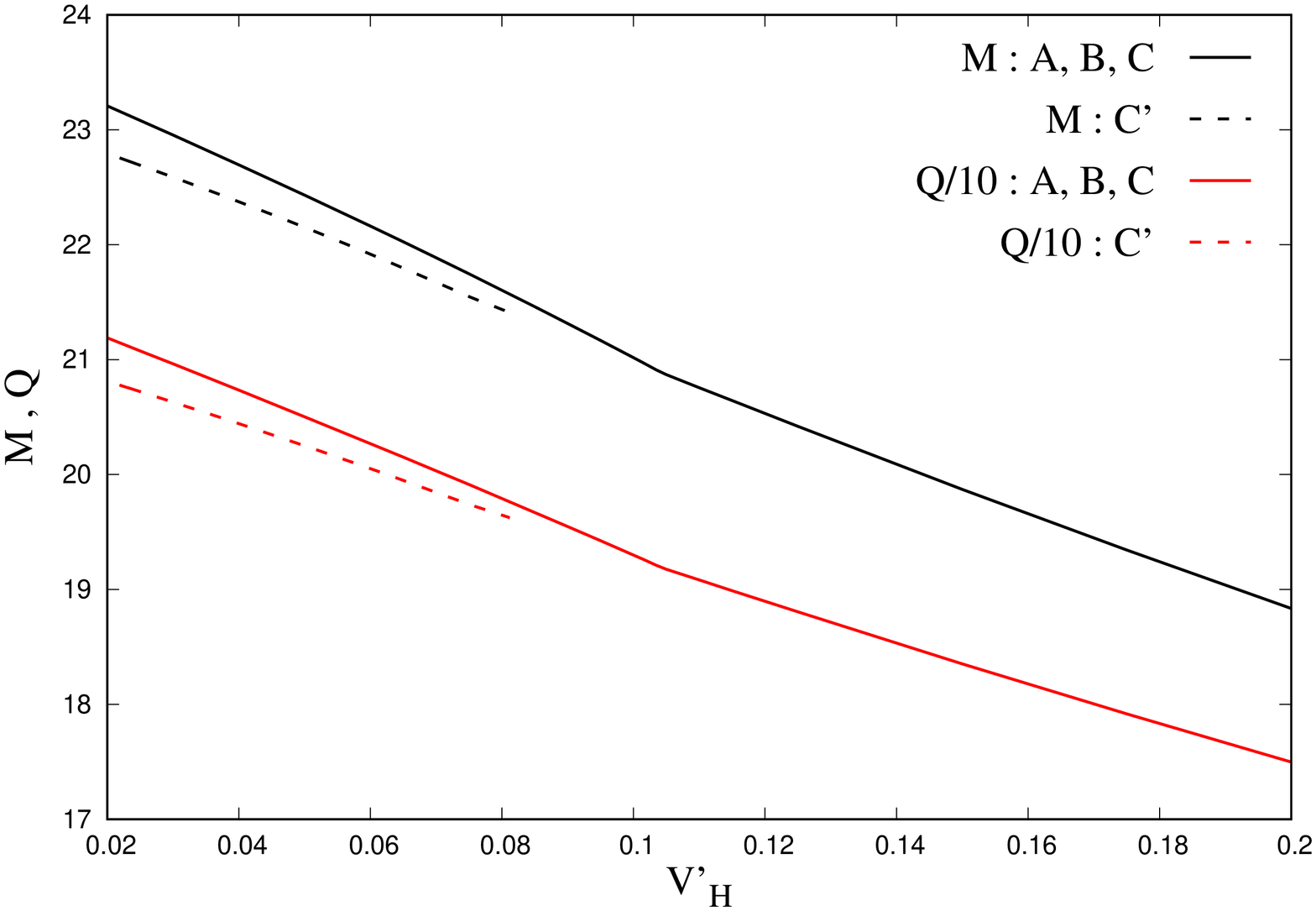}}
\caption{{\it Left}: We show the dependence of the mass $M$ and electric charge  $Q$ of black holes with scalar hair on the value of the electric field on the horizon, $V_H\equiv V'(x_h)$, for $\alpha = 0.012$, $e = 0.08$, $x_h=1$.
{\it Right}: More details of the dependence of the mass $M$ and electric charge $Q$ on $V'(x_h)$ for $\alpha = 0.012$, $e = 0.08$, $x_h=1$
for small values of $V'(x_h)$.
\label{fig:data_rh_1_vph}
}
\end{center}
\end{figure}

Interestingly, we find only one branch of solutions when plotting the physical quantities in function of the electric
field at the horizon $V_H\equiv V'(x_h)$, see Fig. \ref{fig:data_rh_1_vph}. Decreasing $V'(x_h)$ we find that the mass $M$ and electric charge
$Q$ continuously increase. 

Very similar to what has been discussed for boson stars above 
and had already been presented briefly for black holes in \cite{Brihaye:2021phs}, the solutions on branch C develop spatial oscillations in the scalar field function. Again, this is related
to the fact that the effective mass $m_{\rm eff}^2$ of the scalar field (see (\ref{eq:effective_mass})) becomes negative in a given
interval of $x$ due to the metric function $N(x)$
being close to, but actually never becoming equal to zero. 

A new feature in comparison to \cite{Brihaye:2021phs} is that we find not only one branch of solutions with spatial oscillations, but several branches. This is indicated in Fig.\ref{fig:data_rh_1} and Fig.\ref{fig:data_rh_1_vph} by branch C$^{\prime}$. \footnote{In fact, the existence of several branches of solutions with very similar physical properties in this parameter regime  makes the numerical analysis very tedious.}

\begin{figure}[ht!]
\begin{center}
{\includegraphics[width=5cm, angle=-90]{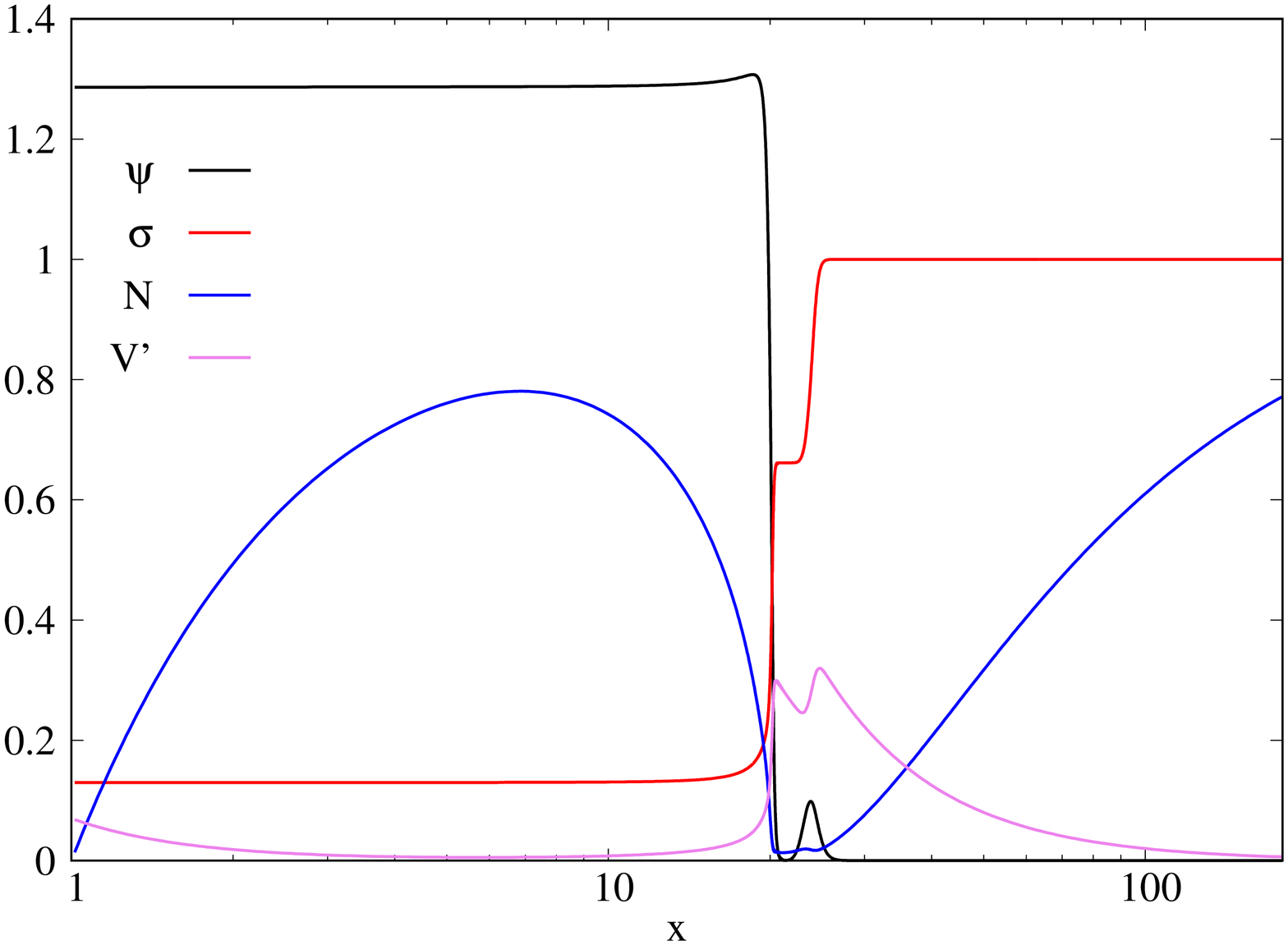}}
{\includegraphics[width=5cm,angle=-90]{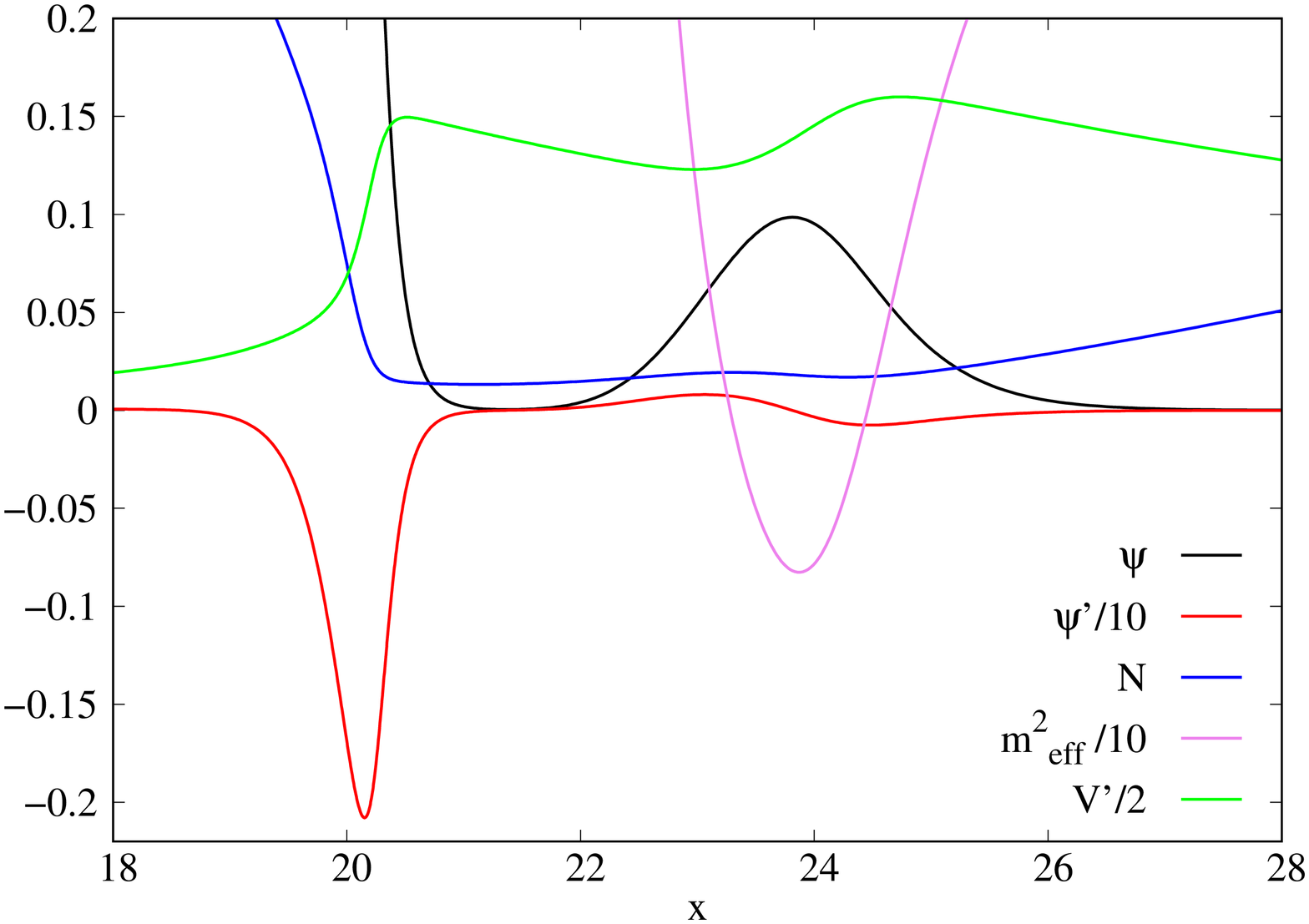}}
\caption{{\it Left}: We show the profiles of the metric functions $N$ and $\sigma$, the scalar field function $\psi$, and the electric field function $V'$ for a black hole solution with $\alpha = 0.012$, $e = 0.08$,$x_h=1$ and $V'(x_h) = 0.07$.  This is a solution on branch C with $Q \approx 200$.
{\it Right}: The details of the solution close to $x \sim 24$.
\label{fig:profile_rh_1_C}
}
\end{center}
\end{figure}

\begin{figure}[ht!]
\begin{center}
{\includegraphics[width=5cm,angle=-90]{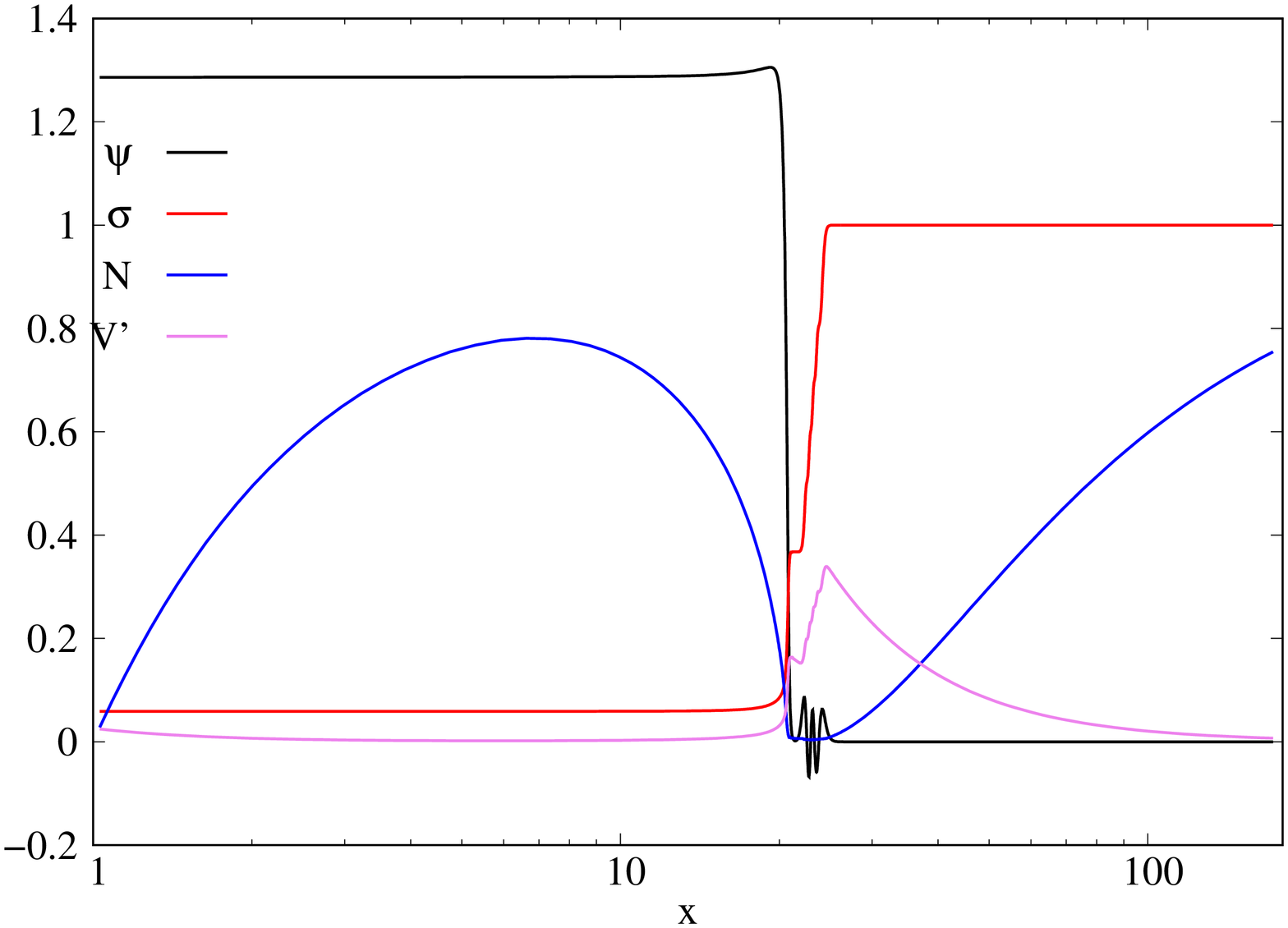}}
{\label{s0}\includegraphics[width=5cm,angle=-90]{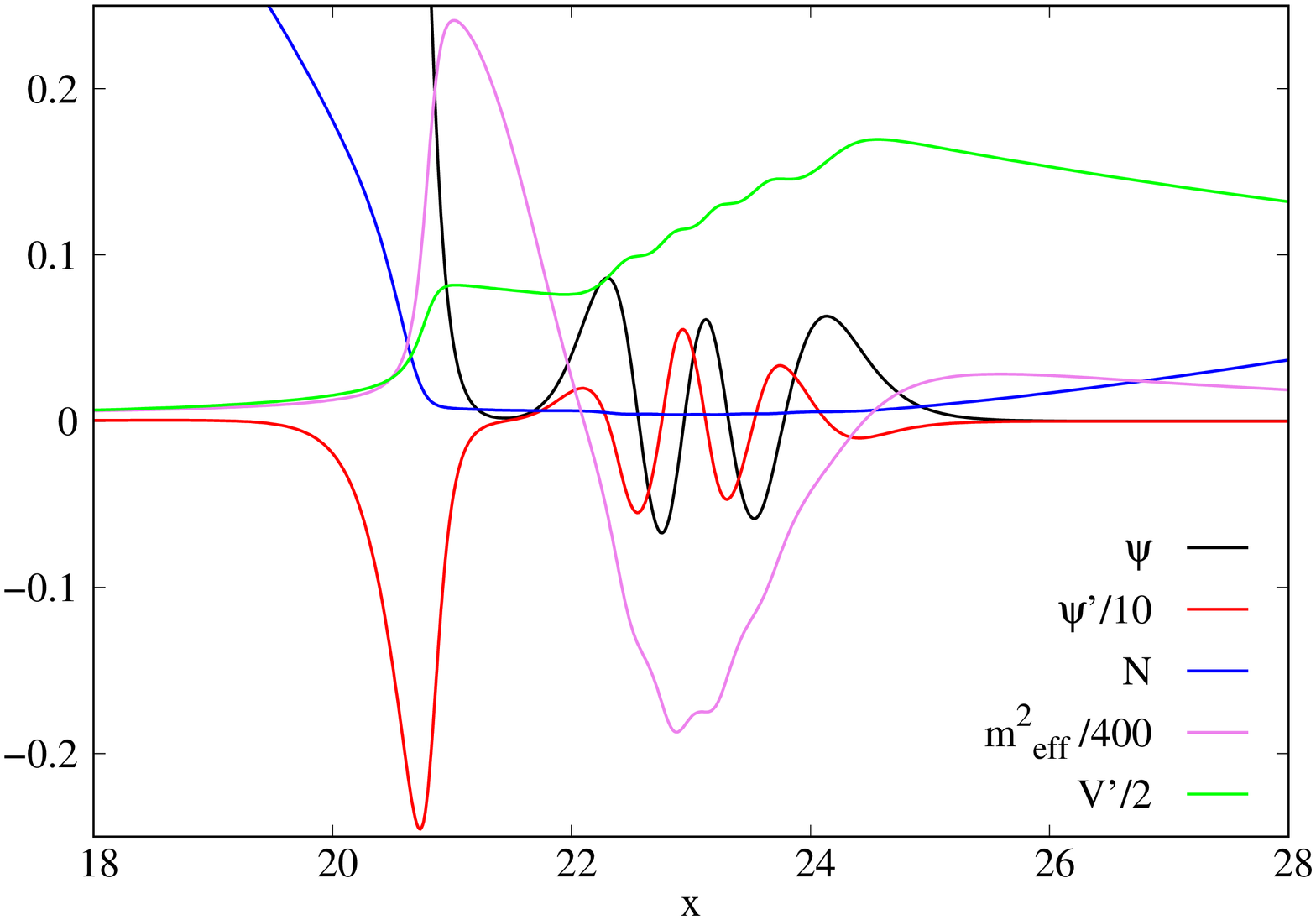}}
\caption{{\it Left}: 
We show the profiles of the metric functions $N$ and $\sigma$, the scalar field function $\psi$, and the electric field function $V'$ for a black hole solution with $\alpha = 0.012$, $e = 0.08$,$x_h=1$ and $V'(x_h) = 0.07$.  This is a solution on branch C$^{\prime}$ with $Q \approx 207$.
{\it Right}: The details of the solution close to $x \sim 24$.
\label{fig:profile_rh_1_Cp}
}
\end{center}
\end{figure}

In order to explain the difference (and origin) of these different branches, we show a solution on branch C in Fig.\ref{fig:profile_rh_1_C} and a solution on branch C$^{\prime}$ in
Fig.\ref{fig:profile_rh_1_Cp}. These solutions
differ in the value of the electric charge $Q$ and $M$.
We find that the solutions on branch C$^{\prime}$ have lower mass $M$ and charge $Q$ for fixed value of the electric field on the horizon
$V'(x_h)$ as compared to the solutions on branch $C$.
In fact, Fig. \ref{fig:profile_rh_1_C} and
Fig.\ref{fig:profile_rh_1_Cp} show that the solutions on branch C$^{\prime}$
have fewer spatial oscillations in comparison to those on branch C.

We believe that the argument about the negativity of 
the effective mass $m^2_{eff}$ allowing spatially oscillating scalar field solutions also suggests that several solutions corresponding to a different number of spatial oscillations in the scalar field should
exist. We have managed to construct two examples for this case
under the requirement of sufficiently small numerical error (typically a relative error on the order of $10^{-6}$). 
The solution on branch C (see Fig.\ref{fig:profile_rh_1_C}) show only one local maximum in the region where $m^2_{eff} < 0$, while the
solution on branch C$^{\prime}$ possesses three local maxima. Our results indicate that the mass $M$ and charge $Q$ increase (slightly) with the increase
in number of oscillations. 

\subsection{Dependence on horizon area}
Considering now the dependence of the results presented above
on the area of the horizon of the black hole, we have followed the approach used in  \cite{Herdeiro:2020xmb} and solved the equations for a fixed value of the asymptotic value of the electric field potential $\Phi$ and
varied the horizon radius $x_h$. We shows values of $\Phi$ for which branches $A$ and $B$ coexist. Our results show that two branches of black holes solutions exist when varying $x_h$. 
They merge at a maximal value of $x_h=x_{h,{\rm max}}$, while for
$x_h\rightarrow 0$ the black hole solutions approach the boson star solutions. In this limit, the temperature of the solution diverges
due to the fact that the metric function $N(x)$ develops an infinite derivative at $x_h$ when $x_h\rightarrow 0$. This is connected to the fact that the boundary conditions for the black holes at $x_h$ is $N(x_h)=0$, while $N(x=0)=1$ for the boson stars. Note that this is hence an artifact of the boundary conditions. 

\begin{figure}[ht!]
\begin{center}
{\includegraphics[width=5cm, angle=-90]{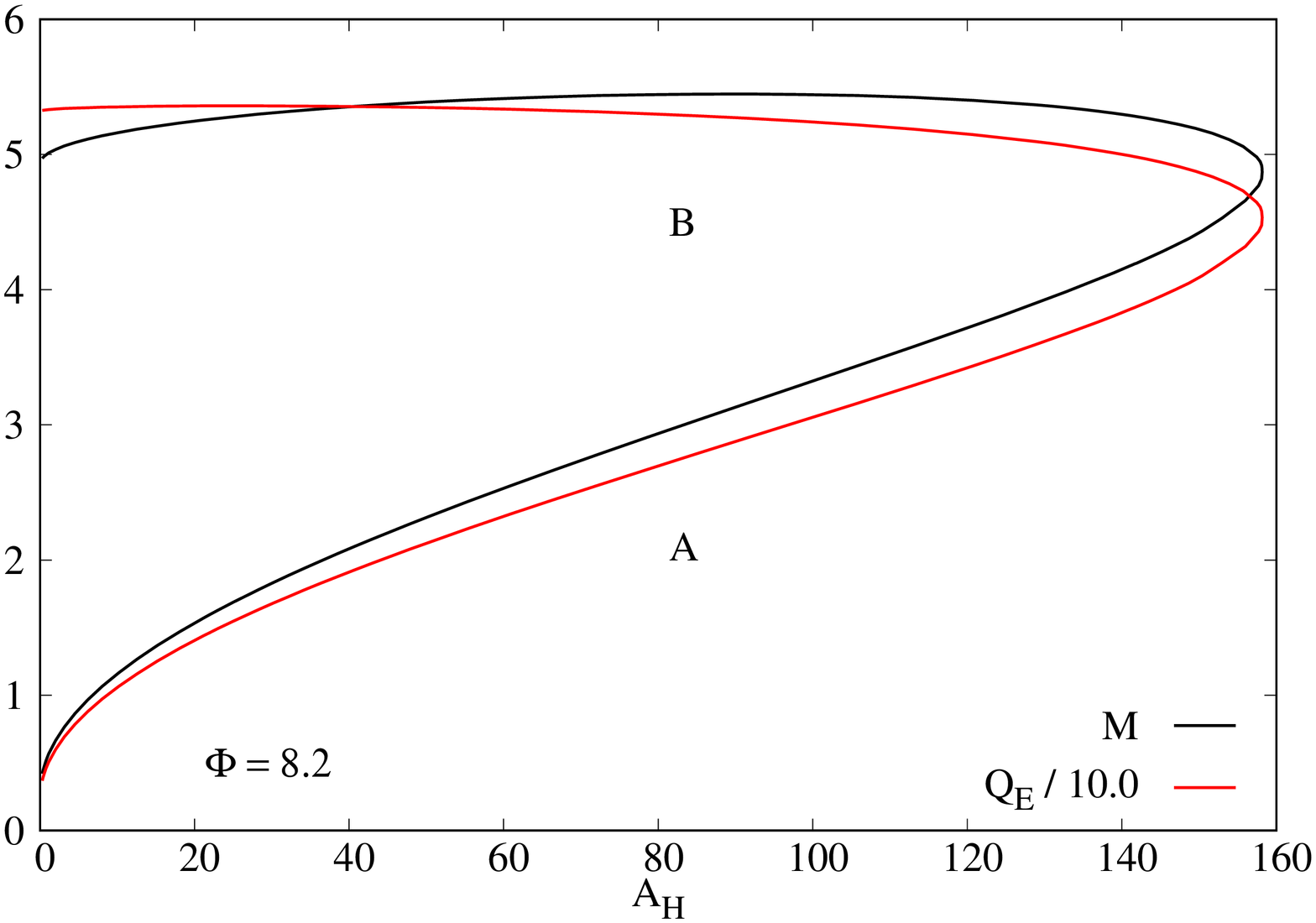}}
{\includegraphics[width=5cm,angle=-90]{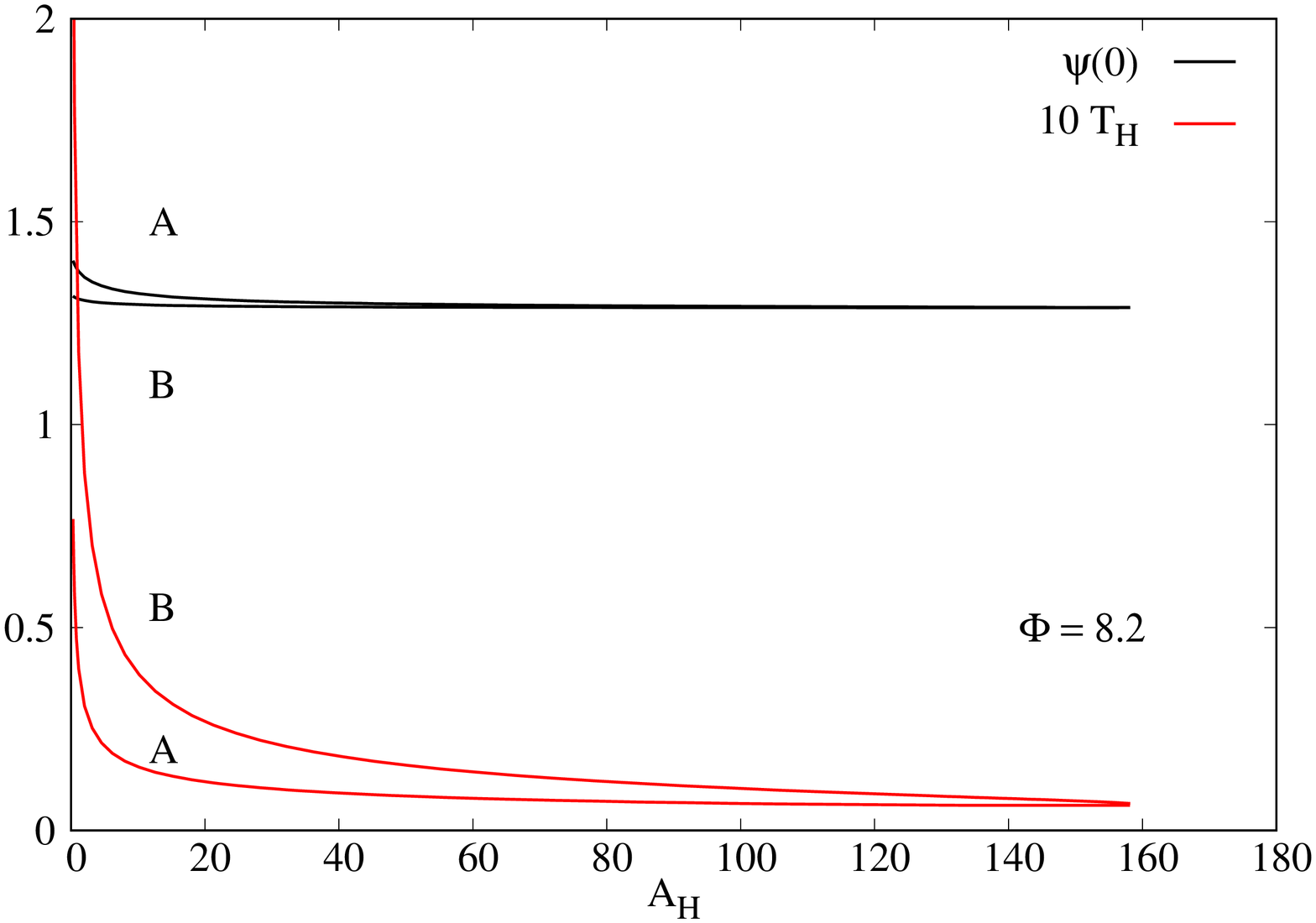}}
\caption{{\it Left}: We show the dependence of the mass $M$ and electric charge $Q$ 
on the value of the horizon area $A_H=4\pi x_h^2$ for black holes with $\Phi = 8.2$.
{\it Right}: We show the dependence of the temperature $T_H$ on the value of the scalar field on the horizon $\psi(x_h)$.
\label{fig:data_rh_vary}
}
\end{center}
\end{figure}

An example of such a family is given in Fig. \ref{fig:data_rh_vary} for  $\Phi = 8.2$. The values  $M$, $Q$ (left) and $\psi(x_h)$, $T_H$ (right) are given
in function of the horizon area $A_H= 4 \pi r_h^2$. The solutions on branch A have lower temperature $T_H$ and higher values of $\psi(x_h)$ in comparison to those on branch B. For the branches C and C$^{\prime}$, we find that when increasing $x_h$ the temperature $T_H$ decreases.

In order to clarify some of the features of the new solutions found,  we have studied some thermodynamical properties of the solutions by introducing the following reduced quantities~:
\be
\label{eq:xyz}
        X \equiv \frac{\sqrt{4 \pi G} Q}{M} \ \ , \ \   
				Y \equiv \ \ \frac{A_H}{16 \pi M^2}  \ \ , \ \ 	
				Z \equiv   8 \pi  M T_H  \ \ .
\ee
These have been presented for several solutions in \cite{Herdeiro:2020xmb} and we extend these results here by adding the new branches of solutions
that we have found  in \cite{Brihaye:2021phs} and in this present paper. Note that for the RN solution we have~:
\be
     T_H = \frac{1}{4 \pi r_h} (1-W) \ \ , \ \ M = \frac{r_h}{2}(1+ W) \ \ , \ \ W \equiv \frac{4 \pi G Q^2}{r_h}
\ee
such that  the quantities given in (\ref{eq:xyz}) read~:
\be
      X =  \frac{2 \sqrt{W}}{1+W} \ \ , \ \   
				Y = \frac{1}{(1+W)^2} \ \ , \ \ 	
				Z  = (1-W)(1+W) \ \ , \ \   {\rm with} \ \ W \equiv \frac{4 \pi G Q^2}{r_h}  . 
\ee
with $W \equiv \frac{4 \pi G Q^2}{r_h} $.

\begin{figure}[h!]
\begin{center}
{\includegraphics[width=5cm, angle=-90]{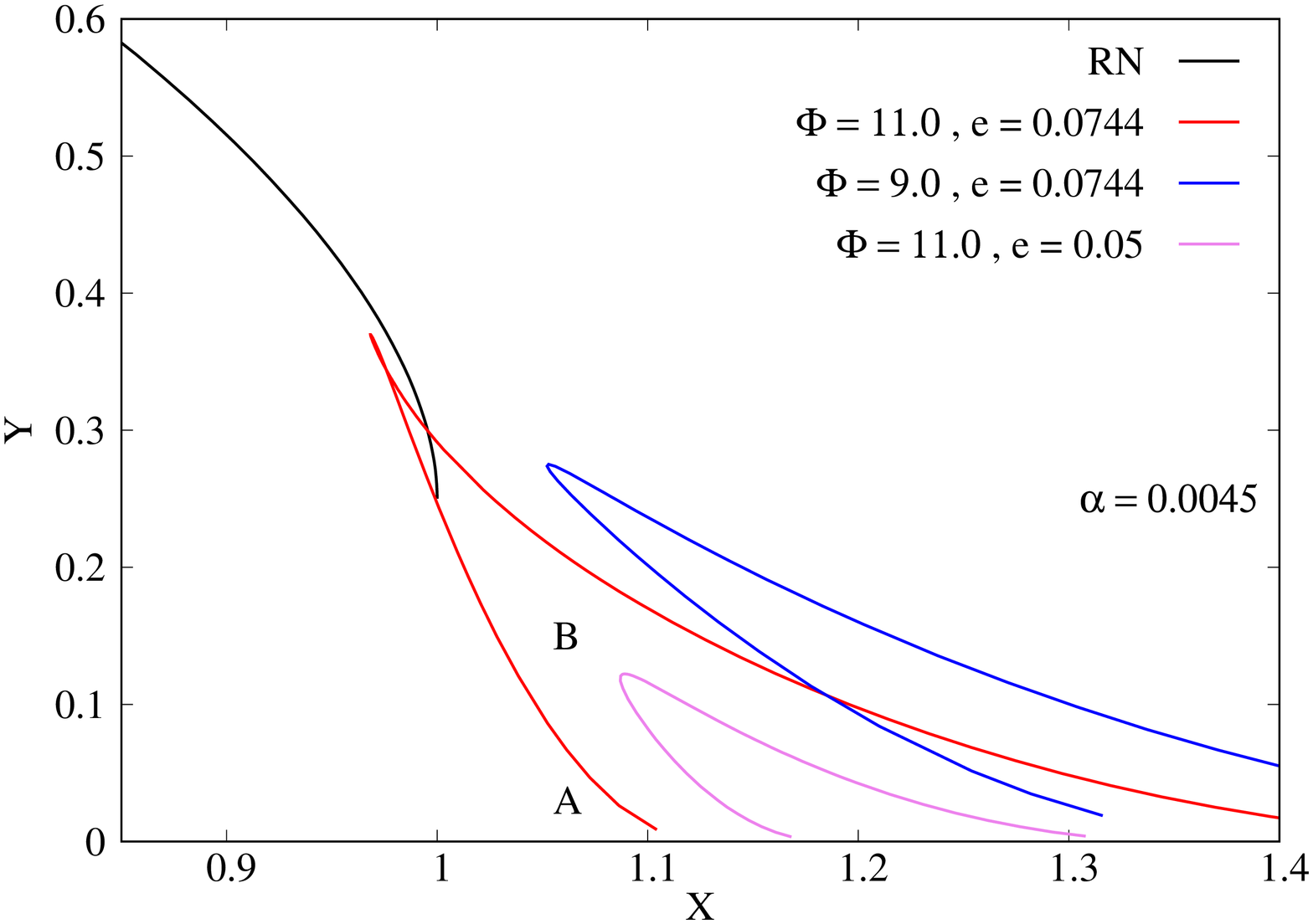}}
{\includegraphics[width=5cm,angle=-90]{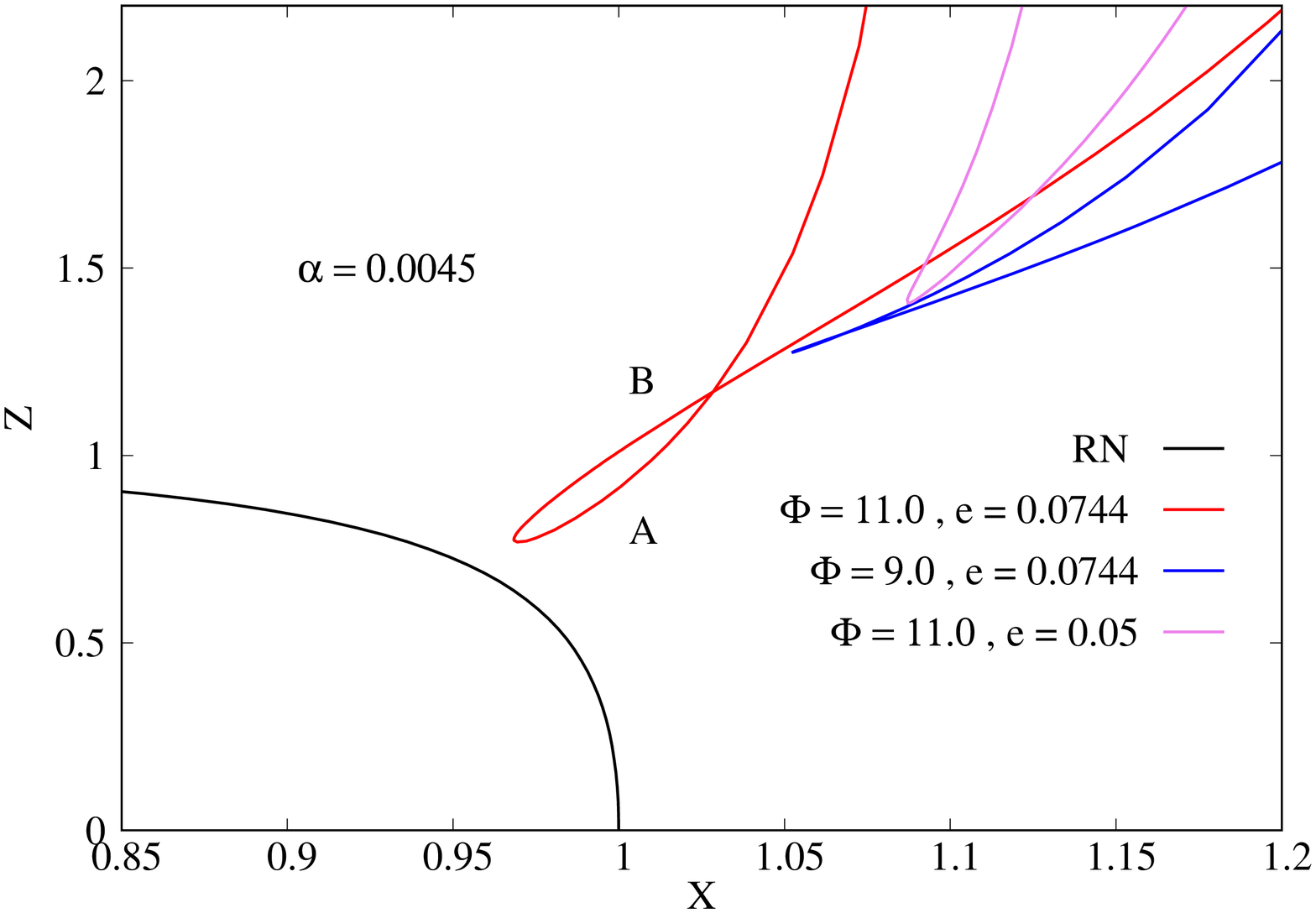}}
{\caption{{\it Left}: The $X$-$Y$ dependence for black holes with standard scalar hair for  
three different choices  of ($e$,$\Phi$) and $\alpha = 0.0045$. Note that here the horizon radius $x_h$ is varied along the branches. 
{\it Right}: The corresponding $X$-$Z$ dependence. }}
\end{center}
\label{NEWXYZ1}
\end{figure}

In  Fig. 15 we give the $X-Y$ and $X-Z$ plots, respectively, 
for $\alpha = 0.0045$ and for several values of $e$ and $\Phi$.
The black line corresponds to the RN solutions, while the red one corresponds to $e=0.0724$, $\Phi=11$. 
This latter choice of parameters has first been presented first in \cite{Herdeiro:2020xmb} and is used here as a means to understand the 
dependence of the quantities $X$, $Y$, $Z$ on the parameters $e$ and $\Phi$, respectively. We have chosen
 $e = 0.05$, $\Phi = 11$  (magenta) and $e=0.0724$, $\Phi =9$ (blue), which we believe exemplify the dependence well. 
More precisely, we find that 
\begin{itemize}
\item when we plot quantities versus $X$ (the charge to mass ratio), two branches of black hole solutions exist. They
are distinguished by different values of $Y$ and the branches join  at a minimal value of $X$,
\item the limit $r_h \to 0$  corresponds to an increase of the quantity $X$,
\item the temperature (represented by the quantity $Z$) stays positive for all black holes with scalar hair and $Z$ increases
       for decreasing horizon radius and is also bigger than the corresponding value for the RN solution at the same value of $X$. 
\end{itemize}

For comparison, we have then chosen the coupling constants equal to those in \cite{Brihaye:2021phs}
and studied the $X$-$Y$ and $X$-$Z$-dependence, respectively, for the black holes with wavy scalar hair. 
These new results are shown in Fig. 16 for $\alpha=0.012$, $e=0.08$ and several values of $\Phi$. Clearly, the behaviour
is qualitatively very different to that shown in Fig. 15. In particular, we note the following~:
\begin{itemize}
\item only one solution exists for a fixed value of charge to mass ratio $X$,
\item the interval in $X$ for which solutions exist increases with $\Phi$,
\item for small enough values of $\Phi$ the solutions of branches $A$ and $B$ are essentially ''overcharged'' in a sense
that the charge to mass ratio exceeds $X=1$ corresponding to the extremal RN case,
\item in the case $\Phi=8.6$ the $X$-$Y$ curve shows a pronounced maximum close to $X = 1$. 
Note that this curve  stops at $X \sim 0.87$ due to  numerical difficulties
to construct solutions with very small horizon. 
\end{itemize}

\begin{figure}[h]
\begin{center}
{\includegraphics[width=5cm, angle=-90]{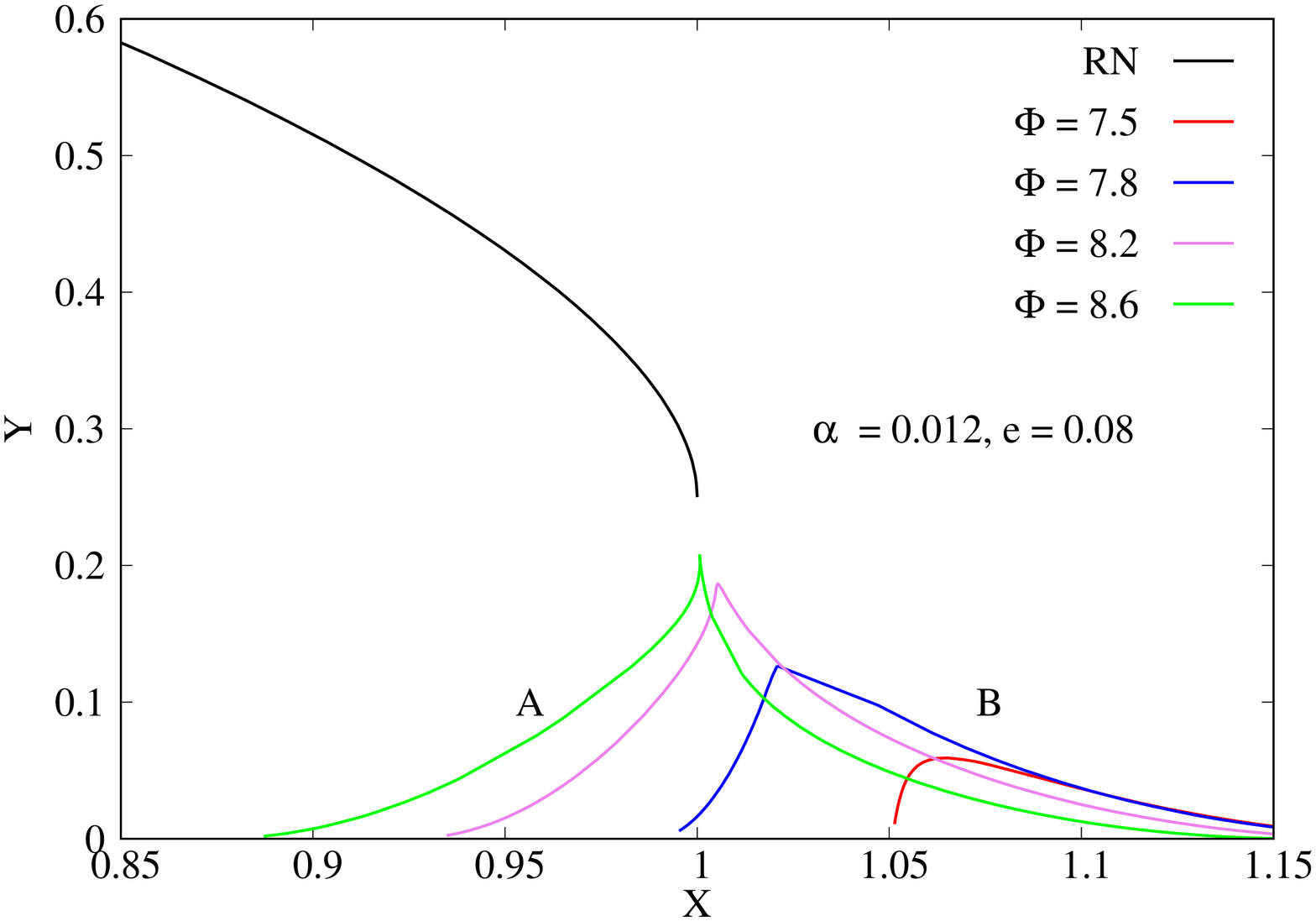}}
{\includegraphics[width=5cm,angle=-90]{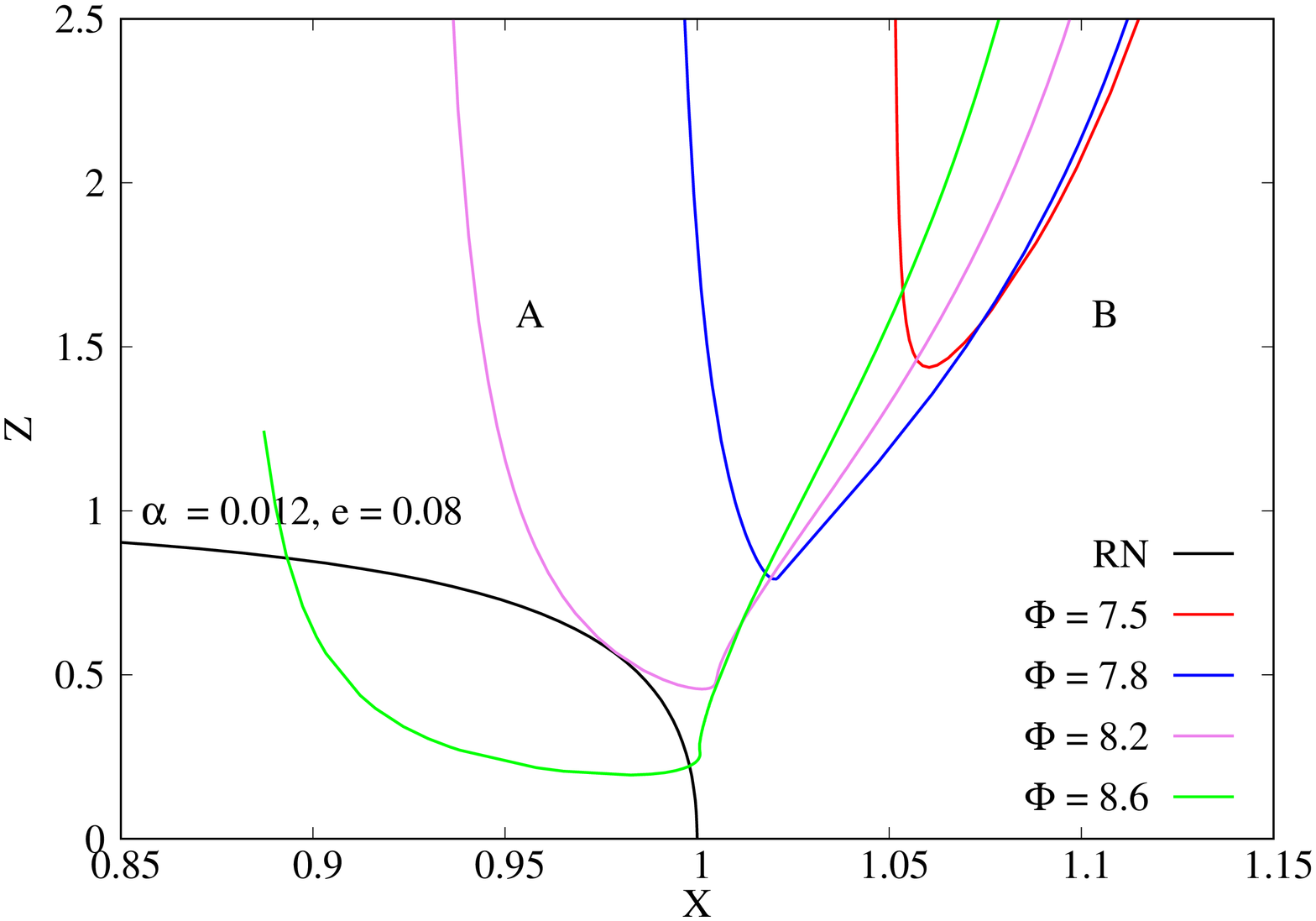}}
{\caption{{\it Left}: The $X$-$Y$ dependence of black holes with wavy scalar hair for  
 $e=0.08$, $\alpha = 0.012$ and several values of $\Phi$. 
{\it Right}: The $X$-$Z$ dependence for the same solutions.}}
\label{NEWXYZ2}
\end{center}
\end{figure}
\subsection{Black hole interior}  
As a final study, we have investigated the interior of these new type of black hole solution. In fact, 
the knowledge of the solution for $x \in [x_h, \infty)$ allows us to integrate the equations for $x< x_h$
by using the (numerical) values of the fields at $x=x_h$ as  boundary conditions. We have studied a few cases and found that
the physical singularity is approached at a value of the radial coordinate $x=x_s$ with $0 < x_s < x_h$. In particular the
geometric invarariants such as the Ricci and Kretschmann scalar, respectively, diverge for $x \to x_s$.
Our numerical analysis reveals further that $x_s$ decreases when moving along the branch $A$ to branch $B$ and $C$ such that for the solutions
on branch $C$ we find the curvature singularity being located very close to $x=0$.  Moreover, the value of the scalar field function is pratically constant in this interval, i.e. $\psi(x) \sim \psi(x_h)$ 
for $x< x_h$. 

\section{Geodesic motion in the presence of wavy scalar hair}
In order to demonstrate that the new type of scalar hair we have discussed here can lead to observable consequences, we discuss briefly the behaviour of test particles in the space-time of the black holes. Similar arguments appear for boson stars and a detailed analysis will be presented elsewhere.

Using the symmetries of the space-time and denoting the energy of the test particle by $E$ and its angular momentum by $L_z$, the equation describing geodesic motion 
$g_{\mu\nu}\dot{x}^{\mu} \dot{x}^{\nu}=\varepsilon$ takes the form
\begin{equation}
\label{eq:geodesic}
\sigma^2\dot{x}^2 + V_{\rm eff}(x)= E^2 \ \ , \ \     
V_{\rm eff}(x)=
N\sigma^2 \left(\frac{L_z^2}{x^2} - \varepsilon\right) \ , 
\end{equation}
where the dot denotes the derivative with respect to an affine parameter. Moreover, $\varepsilon$ takes on the value $0$ for massless particles and 
$-1$ for massive particles, respectively. 

\begin{figure}[ht!]
\begin{center}
{\includegraphics[width=7cm,angle=-90]{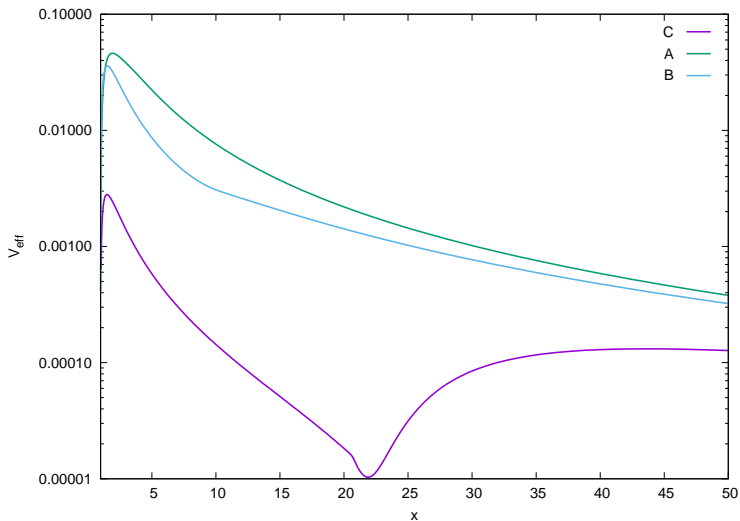}}
\caption{We show the effective potential $V_{\rm eff}$ (see (\ref{eq:geodesic})) for a massless test particle
with $L_z=1$ in the space-time of three different black holes with scalar hair corresponding to branches A, B and C. For all three space-times we have chosen $x_h=1$, $\alpha=0.012$ and $e=0.08$. The black hole on branch A has $\Phi=8.2$, $Q=11.6$, the one on branch B has $\Phi=8.2$, $Q=53.5$,
while the black hole on branch C has $\Phi=9.84$, $Q=200$, respectively. 
The minimum of the effective potential for the case $C$ appears at $r\approx 21.8$.
}
\label{fig:veff_geodesic}
\end{center}
\end{figure}

We show the effective potential $V_{\rm eff}$ for a massless particle in some black hole space-times in Fig. \ref{fig:veff_geodesic}. Note that for
$\varepsilon=0$, the numerical value of the angular momentum $L_z$ does not influence the qualitative discussion in the following - that is why we have set it to unity for this plot. 
As is obvious, there exists a new feature in the space-time of a black hole with wavy scalar hair (solution C) as compared to 
those with ``standard'' scalar hair: namely a local minimum in the exterior space-time of the black hole. For the example plotted, 
the local minimum is located at $x\approx 21.8$ and has value $\approx 10^{-5}$.
Hence, a massless particle with $E^2/L_z^2\approx 10^{-5}$ would move in a {\it stable circular orbit} around the black hole that carries wavy scalar hair. This suggests that these type of black holes would
possess a stable photon sphere well outside their horizon.

\section{Conclusions and Outlook}
In the era of multi-messenger astrophysics and new telescopes and experiments being proposed to 
address long-standing questions in modeling all four interactions satisfactorily, extensions of the best model that we have to date for the gravitational
interaction have been proposed. Often this involves new direct couplings between matter fields and the space-time curvature.
However, even in minimally coupled models, i.e. in models where General Relativiy is coupled minimally to matter fields interesting new 
phenomena appear. In particular, the inclusion of scalar fields has been proven to be very fruitful in this context and
a number of no-hair conjectures can be violated under certain conditions (see e.g. \cite{Herdeiro:2015waa} for a review). 
The interplay between electromagnetic and scalar fields in curved space-time leads to new interesting solutions such as globally
regular objects, so-called {\it boson stars} \cite{kaup,misch,flp, jetzler,new1} that are (still) a viable alternative to black holes as they can be as heavy and as large as the latter. 

In this paper we have demonstrated that new features appear in a complex scalar field, U(1) gauged model when the curvature of space-time
is strong and the scalar field is self-interacting. It is well-known that black holes can carry scalar hair in this model, however, we find that
the form of this scalar hair is more subtle than previously thought. We find that for large gravitational coupling and 
large electric charge the solutions develop spatial oscillations in the scalar field, hence a new form of scalar hair that we have denoted {\it wavy scalar hair}. 
These oscillations appear in a region of space-time where $g^{rr}$ is very close to zero on an extended interval of the radial coordinate, but actually
never is zero. This leads to the effective mass of the scalar field (as defined via a linearization of the scalar field equation) becoming negative
and the scalar field equation has the form of a harmonic oscillator equation. 
Interestingly, these oscillations also appear when considering boson stars in this model, i.e. the existence of an event horizon in the space-time is not crucial.
This argumentation suggests - and has been confirmed numerically by us - that a whole discrete tower of solutions should exist. 

The observational consequences can only be speculated on at this moment, but we have demonstrated that massless (and, in fact, also massive)
test particles would be able to move on stable circular orbits around black holes and boson stars, respectively. This observation suggests that black holes
and boson stars should possess a photon sphere. It will be interesting to investigate these observational consequences further, in particular with view to
the results given in \cite{new_citations}.

An interesting extension of our results would be to study whether the new features we observe are generic, i.e. exist e.g. also in higher dimensions
and/or for rotating and uncharged solutions.

\clearpage


 \end{document}